\documentclass[aps,twocolumn,showpacs,preprintnumbers,nofootinbib,prd,superscriptaddress,groupedaddress,10pt]{revtex4-1}

\makeatletter
\def\l@subsubsection#1#2{}
\def\l@subsubsubsection#1#2{}
\makeatother

\usepackage{graphicx,amssymb,amsmath,amsthm,amsfonts,epsfig,epsf}
\usepackage[usenames]{color}
\usepackage{epstopdf}

\usepackage{aas_macros}
\usepackage{bm}
\usepackage{dcolumn}
\usepackage{latexsym}
\usepackage{rotating}
\usepackage{longtable}

\usepackage{enumerate}
\usepackage{tensor,multirow}
\usepackage{url}
\usepackage[linktocpage]{hyperref}

\def\be{\begin{equation}}
\def\ee{\end{equation}}
\def\beq{\begin{eqnarray}}
\def\eeq{\end{eqnarray}}
\begin{document}

\title{Tidal effects and disruption in superradiant clouds: a numerical investigation}
\author{Vitor Cardoso}
\affiliation{CENTRA, Departamento de F\'{\i}sica, Instituto Superior T\'ecnico -- IST, Universidade de Lisboa -- UL,
Avenida Rovisco Pais 1, 1049 Lisboa, Portugal}
% 
%\affiliation{Institute for Theoretical Physics, University of Amsterdam, PO Box 94485, 1090GL, Amsterdam, The Netherlands}
%
\author{Francisco Duque}
\affiliation{CENTRA, Departamento de F\'{\i}sica, Instituto Superior T\'ecnico -- IST, Universidade de Lisboa -- UL,
Avenida Rovisco Pais 1, 1049 Lisboa, Portugal}
\author{Taishi Ikeda}
\affiliation{CENTRA, Departamento de F\'{\i}sica, Instituto Superior T\'ecnico -- IST, Universidade de Lisboa -- UL,
Avenida Rovisco Pais 1, 1049 Lisboa, Portugal}

\begin{abstract} 
The existence of light, fundamental bosonic fields is an attractive possibility that can be tested via black hole observations. 
We study the effect of a tidal field -- caused by a companion star or black hole -- on the evolution of superradiant scalar-field states around spinning black holes. For small tidal fields, the superradiant ``cloud'' puffs up by transitioning to excited states
and acquires a new spatial distribution through transitions to higher multipoles, establishing new equilibrium configurations.
For large tidal fields the scalar condensates are disrupted; we determine numerically the critical tidal moments for this to happen and find good agreement with Newtonian estimates.
We show that the impact of tides can be relevant for known black-hole systems such as the one at the center of our galaxy or the Cygnus X-1 system. The companion of Cygnus X-1, for example, will disrupt possible scalar structures around the BH
for gravitational couplings as large as $M\mu\sim 2\times 10^{-3}$.
%The S2 star, for example, will disrupt possible scalar structures around the black hole at the center of our galaxy, for gravitational %couplings as large as $M\mu=0.009$, with $M$ the mass of the supermassive black hole and $\mu$ the mass parameter of the scalar field.
%
\end{abstract}

%%%%
%%%%
\maketitle

%%%%%%%%%%%%%%%%%%%%%%%%%%%%%%
\section{Introduction}
%%%%%%%%%%%%%%%%%%%%%%%%%%%%%%
The matter content of our universe is largely unknown, and has been the focus of an incredible effort in the last decades~\cite{Bertone:2018xtm}. Most experimental searches are based on putative couplings between dark matter (DM) and standard model fields. The null results of such experiments provide interesting constraints on the strength of such couplings, but are otherwise unable to shed light on new fundamental constituents of the universe.

The universal nature of gravity suggests that new fields or particles behave in the same way as their standard model cousins when placed in gravitational fields. It is no surprise therefore that the only but solid evidence for DM interaction is so far of a purely gravitational nature. The advent of gravitational-wave (GW) astronomy provides a compelling case to understand further the behavior of DM in strong gravity situations~\cite{Barack:2018yly,Baibhav:2019rsa,Maggiore:2019uih}. Of particular relevance in this context are black hole (BH) spacetimes. In vacuum general relativity these are the simplest macroscopic object one can conceive of, and ideal to be used as testing grounds for the presence of new fields or extensions of general relativity~\cite{Bertone:2018xtm,Barack:2018yly,Baibhav:2019rsa,Maggiore:2019uih,Cardoso:2016ryw,Brito:2015oca}.

The DM density in our universe is measured to be small enough that its effects on the dynamics of compact objects -- BHs in particular -- are perturbatively small. The imprint that DM leaves on the GW is correspondingly small, but potentially measurable by future GW detectors~\cite{Eda:2014kra,Macedo:2013qea,Barausse:2014tra,Cardoso:2019rou}. However, should ultralight bosonic degrees of freedom exist in nature~\cite{Arvanitaki:2009fg,Marsh:2015xka}, superradiance will give rise to the development of massive structures (``clouds'') around spinning astrophysical BHs. 
This is a general mechanism that requires only minimal ingredients. A simple minimally coupled massive field with no initial abundance
suffices: a superradiant instability sets in, extracting rotational energy away from the BH and depositing it in any small boson fluctuation outside the BH. For the mechanism to be effective, the BH radius $2GM/c^2$ needs to be of the order of the boson Compton wavelength $G/(c^2 \mu)$ for a particle of mass $m_B$ (here $\mu=Gm_B/(c\hbar)$ is the mass parameter that will appear in all our equations). In other words, the mechanism is effective when $M\mu\sim 1$. However, because BHs in our universe appear in a wide range of masses -- that vary over 8 or more orders of magnitude, superradiance allows to effectively study or rule out boson masses varying by correspondingly large orders of magnitude~\cite{Arvanitaki:2010sy,Brito:2014wla,Brito:2015oca}. 
 
The existence of superradiant clouds would lead to observable signatures, such as peculiar holes in the mass-spin plane of BHs~\cite{Arvanitaki:2010sy,Brito:2014wla}, to monochromatic emission of GWs~\cite{Arvanitaki:2016qwi,Brito:2014wla} and to a significant stochastic background of GWs~\cite{Brito:2017zvb,Brito:2017wnc}. 
The presence of such periodic, non-axisymmetric structures can leave imprints in planetary and stellar orbits, through Lindblad and co-rotation resonances~\cite{Ferreira:2017pth,Boskovic:2018rub}, or -- of interest for GW astronomy -- through floating or sinking orbits~\cite{Cardoso:2011xi,Zhang:2018kib,Zhang:2019eid,Baumann:2019ztm}.
There has been a significant progress in our understanding of the development of the superradiant instabilities~\cite{Brito:2015oca}. There are two main factors that could alter, in a significant way, the formation of heavy boson clouds around BHs. In the presence of couplings between the ultralight boson and standard model fields, for example, the cloud growth can be suppressed, while stimulating bursts of light~\cite{Ikeda:2019fvj,Boskovic:2018lkj}.

Here, we focus instead on the effects that a companion star or BH have on the structure of the boson cloud. Tidal effects were studied recently, at an analytical level, and using Newtonian dynamics for non relativistic fields~\cite{Arvanitaki:2014wva,Zhang:2018kib,Zhang:2019eid,Baumann:2018vus,Berti:2019wnn,Baumann:2019ztm}. The motion of the binary can, at specific orbital frequencies, induce resonant transitions between growing and decaying modes of the boson, that enhance the cloud's depletion and/or transfer energy and angular momentum to the companion through tidal acceleration~\cite{Cardoso:2012zn}. This behaviour would leave distinctive imprints in the GW signal emitted by the binary, both as a monochromatic signal from the cloud or as modifications in the GW waveform of the binary, due to finite-size effects (eg. variations on the spin-induced quadrupole or the tidal Love numbers)~\cite{Baumann:2018vus}. This situation could be of interest for eccentric BH binaries targeted by the space-interferometer LISA~\cite{Berti:2019wnn}.

%%%%%%%%%%%%%%%%%%%%%%%%%%%%%%
\section{Setup}
%%%%%%%%%%%%%%%%%%%%%%%%%%%%%%
Our starting point is that of a Kerr BH spacetime perturbed by a distant companion. The BH
is surrounded by a superradiant cloud, which is assumed to cause negligible backreaction in the spacetime. The geometry will always be kept fixed in this work, in the sense that the scalar field never backreacts back. This working hypothesis holds true for most of the situations of interest~\cite{Brito:2014wla}, and is specially appropriate here: as we explain below the timescales that we can probe
are much shorter than any superradiant-growth timescales.
A companion of mass $M_c$ is now present, at a distance $R$, and located at $\theta=\theta_c,\phi=\phi_c$ in the BH sky. The companion induces a change $\delta ds_{\rm tidal}^2$ in the geometry. Thus, our spacetime geometry is described by
\be
ds^2=ds^2_{\rm Kerr}+\delta ds_{\rm tidal}^2\,.
\ee
For the tidal perturbation induced by the companion, we consider the non-spinning approximation, and we find in Regge-Wheeler gauge that
the dominant quadrupole tidal contribution is~\cite{Cardoso:2017cfl,Cardoso:2019upw,Taylor:2008xy}
\beq
\delta ds^2&=&\sum_m r^2{\cal E}_{2m}Y_{2m}(\theta,\phi)(f^2dt^2+dr^2+(r^2-2M^2)d\Omega^2)\nonumber\\
%
%&+& \, \sum_m \frac{{\cal B}_{2m}}{3}r^3f(S_\theta^{lm}(\theta,\phi)dt\,d\theta+S_\varphi^{lm}(\theta,\phi)dt\,d\varphi) , \label{eq:MetricPerturbation}\\
%
{\cal E}_{2m}&=&\frac{8\pi \epsilon}{5 M^2}\,Y^*_{2m}(\theta_c,\phi_c)\,,\label{eq:MetricPerturbation}
%\quad {\cal B}_{2m}= \mathcal{O}\left(\frac{v}{c}\right) \, ,
\eeq
%
%
%\be 
%\left(S_\theta^{lm},S_\varphi^{lm}\right)\equiv\left(-Y^{lm}_{,\varphi}/\sin\theta,\sin\theta \, Y^{lm}_{,\theta}\right) \,.
%\ee
%
where $f=1-2M/r$ and we neglect subdominant magnetic-type contributions and multipoles higher than the quadrupole. For more details see Appendix~\ref{sec:TidesGR}. We introduce a dimensionless tidal parameter
\be
\epsilon=\frac{M_{c}M^{2}}{R^3}\,,
\ee
which measures the strength of the tidal moment. Coordinates are Boyer-Lindquist at large distance. This approximation is not accurate close to the BH horizon, where spin effects change the tidal description. However, for all the parameters considered here the cloud is localized sufficiently far-away that these effects ought to be very small. We will focus exclusively on static tides (or in other words, we consider large separations $R$).
We stress that we are using coordinates adapted to the BH: the companion position should in general be time-dependent, but we focus exclusively on slowly moving companions.

We consider a massive, minimally coupled scalar field $\Phi$ evolving on the above fixed geometry.
The scalar is described by the Klein-Gordon equation
%\begin{widetext}
\be
\left(\nabla^{\mu}\nabla_{\mu} - \mu^{2} \right) \Phi =0\,.\label{eq:MFEoMScalar}
\ee
%\end{widetext}
%
Note that, for zero rotation, distances $r\gg M$ and non-relativistic fields, the Klein-Gordon equation can be expressed as in Eq. (3.4) of Ref.~\cite{Baumann:2018vus}, amenable to a perturbation treatment. Some of the implications are summarized in the Appendix. However, we consider the Klein Gordon in full generality, by evolving it numerically.
To express Eq.~\eqref{eq:MFEoMScalar} as a Cauchy problem, we use the standard 3+1 decomposition of the metric,
\begin{eqnarray}
ds^2=-\alpha^{2}dt^{2}+\gamma_{ij}(dx^{i}+\beta^{i}dt)(dx^{j}+\beta^{j}dt)\,,
\end{eqnarray}
where $\alpha$ is the lapse function, $\beta^{i}$ is a shift vector, and $\gamma_{ij}$ is the 3-metric on spacial hypersurface.
We also introduce the scalar momentum $\Pi$,
\be
\Pi=-n^{\mu}\nabla_{\mu}\Phi\,.
\ee
The evolution equation for the axion field is written as
\beq
\partial_{t}\Phi&=&-\alpha\Pi+\mathcal{L}_{\beta}\Phi\,,\nonumber \\
\partial_{t}\Pi&=&\alpha(-D^{2}\Phi+\mu^{2}\Phi+K\Pi)-D^{i}\alpha D_{i}\Phi+\mathcal{L}_{\beta}\Pi\,.\nonumber
\eeq
We use Cartesian Kerr-Schild coordinates $(t,x,y,z)$~\cite{Witek:2012tr}.

The quantities extracted from our numerical simulation are multipolar components of the scalar $\Phi$,
\be
\Phi_{l,m}(t,r)=\int d\Omega \Phi(t,r,\theta,\phi) Y_{l,m}(\theta,\phi)\,.
\ee

We use as initial data the following profile, adequate to describing quasi-stationary states around a BH~\cite{Yoshino:2013ofa,Brito:2014wla}
\be
\label{Eq.axion cloud initial data}
\Phi(t,r,\theta,\phi)=A_{0}rM\mu^{2}e^{-rM\mu^{2}/2}\cos(\phi -\omega_{\rm R}t)\sin\theta\,.
\ee
We will also show below that these are indeed good description of stationary states for small couplings $M\mu \lesssim 0.2$.
Here $A_{0}$ is an arbitrary amplitude related to the mass in the axion cloud, and $\omega_{\rm R}\sim \mu$ is the bound-state frequency.

The spacetime of a real astrophysical binary is asymptotically flat. However, because we are using only an approximation to the full problem, where the companion is supposed to be a large distance away, the geometry~\eqref{eq:MetricPerturbation} is no longer asymptotically flat. To avoid unphysical behavior at large distances, we force the geometry to be asymptotically flat, by replacing the far region with
%
%\begin{eqnarray}
%ds^2=\left(1-\mathcal{W}(\frac{r-r_{\rm th}}{w})\right)
%(ds^2_{\rm Kerr}+\delta ds^2_{\rm tidal})+\mathcal{W}(\frac{r-r_{\rm th}}{w})\eta_{\mu\nu},
%\end{eqnarray}
%
%\vc{Why not}
%
\begin{eqnarray}
ds^2=ds^2_{\rm Kerr}+\left(1-\mathcal{W}\right)\delta ds^2_{\rm tidal}\,,
\end{eqnarray}
where $\mathcal{W}=\mathcal{W}(\tilde{r})$ is a following piecewise function
\begin{equation}
\mathcal{W}(\tilde{r})=\left\{
\begin{array}{ll}
1&(\tilde{r}>1)\\
\mathcal{W}_{5}&(0<\tilde{r}<1)\\
0&(\tilde{r}<0).\\
\end{array}
\right.
\end{equation}
Here, $\tilde{r}=(r-r_{\rm th})/w$ and $\mathcal{W}_{5}(\tilde{r})$ is chosen to match smoothly with the required asymptotic behavior, so we choose a 5th-order polynomial satisfying $\mathcal{W}_{5}(1)=1,\mathcal{W}_{5}(0)=\mathcal{W}_{5}'(0)=\mathcal{W}_{5}''(0)=\mathcal{W}_{5}'(1)=\mathcal{W}_{5}''(1)=0$.
The transition region has a width $w=500M$ and is located at $r_{\rm th}/M\simeq \sqrt{0.9\times 5/(8\pi \epsilon)}$.
These parameters were chosen to ensure that the bosonic cloud sits entirely in a region described by Eq.~\eqref{eq:MetricPerturbation}.

The evolution equations were integrated using fourth-order spatial discretization and a Runge-Kutta method.
Accuracy requirements, finite size of the numerical grid and computational power all contribute to limit the timescales that one is able to access. Here, we evolve these systems for timescales $\sim 7000M$.

Although we have results for general BH spin parameter, we focus mostly on states around a {\it non-spinning} BH.
These states are not superradiant in origin, and arise due to the fine-tuned initial data. However, they are extremely long-lived (the decay timescale is of the order of the superradiant growth timescale if the BH was spinning), as we show below, with a lifetime that far exceeds that of all the tidally-induced transitions studied here. Thus, BH spin is important to generate the scalar clouds, but has little impact on {\it some} of the physics of tides. In addition, the tidal field in Eq.~\eqref{eq:MetricPerturbation} is adapted to a non-spinning BH. Our numerical results indeed show only a very mild dependence on BH spin. With the exception of Ref.~\cite{Berti:2019wnn}, all previous results on tidal effects in superradiant clouds focus on the small $M\mu$ coupling parameter, consider a flat background on which the superradiant states evolve, and have only used linearized analysis for small tidal fields. Our framework can go beyond all these limitations.

\begin{figure}[htb]
\begin{tabular}{cc}
\includegraphics[width=8.5cm]{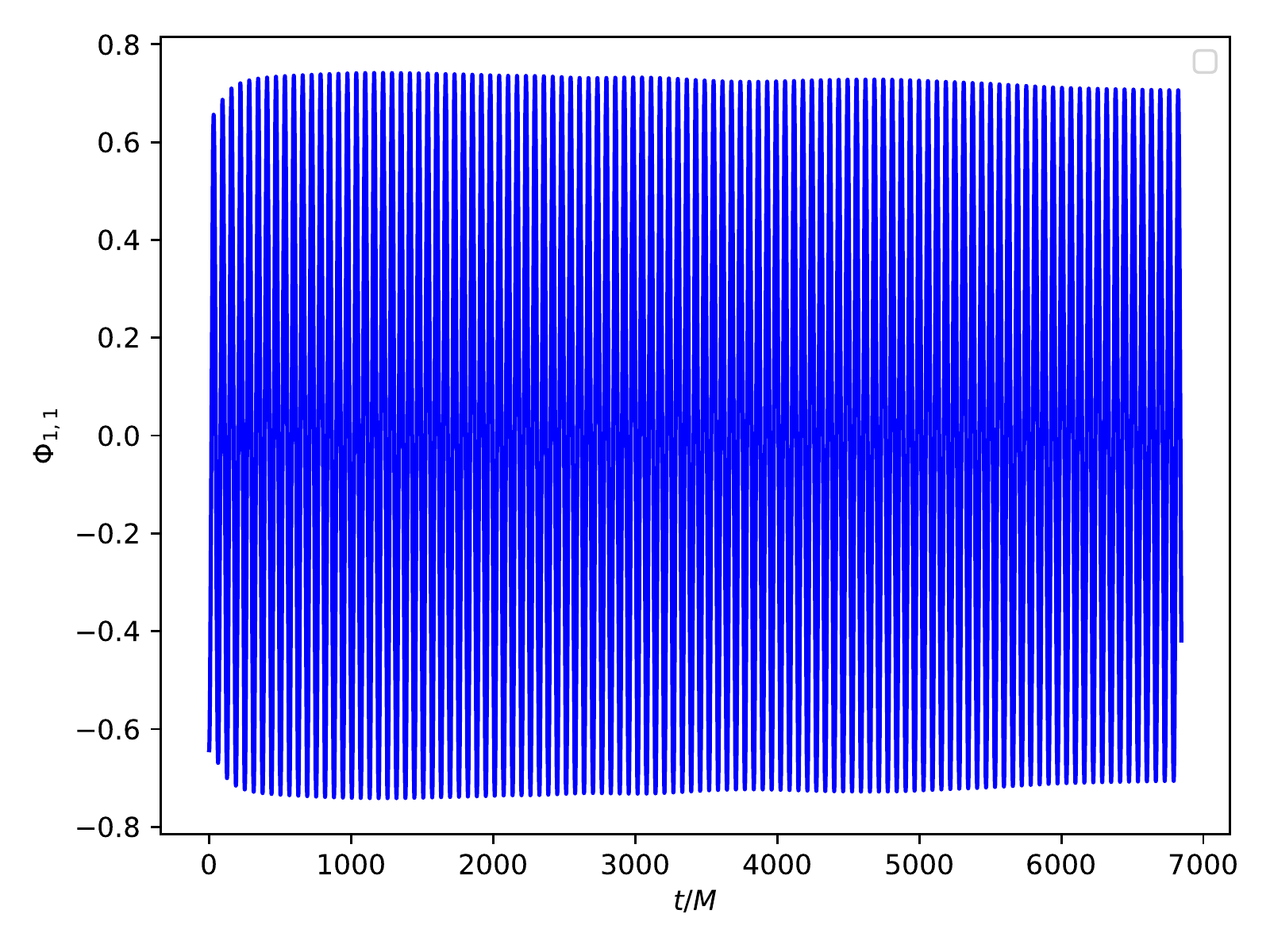}%&\includegraphics[width=8cm]{Plots/Phi_Mc0_a0_mu01.png}
\end{tabular}
\caption{A dipolar scalar cloud around a Schwarzschild BH. This figure shows the time evolution of initial conditions~\eqref{Eq.axion cloud initial data}
for a dipole with gravitational coupling $M\mu=0.1$ around a non-spinning BH, and in absence of a companion ($\epsilon=0$).
The field is extracted at $r=60M$.
\label{fig:a0_Mc0_mu01}}
\end{figure}
\begin{figure}[htb]
\begin{tabular}{cc}
\includegraphics[width=4.3cm]{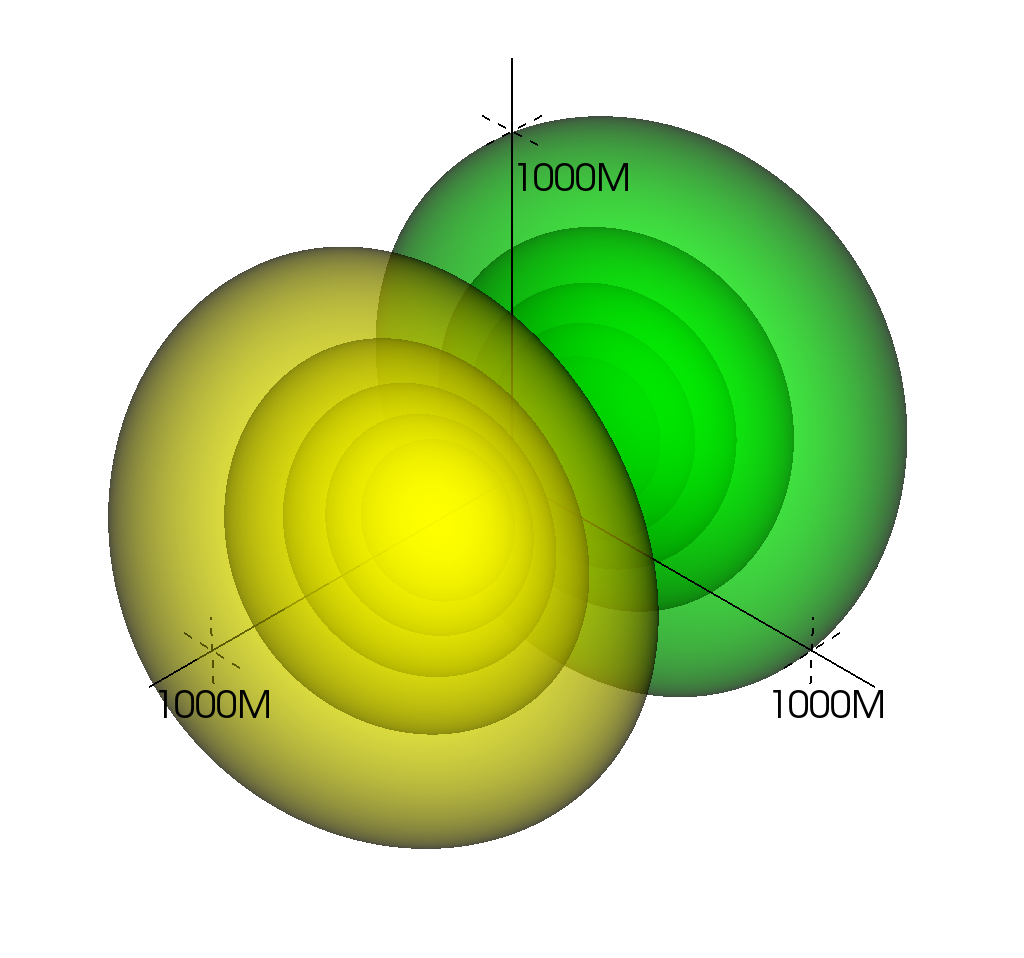}&
\includegraphics[width=4.3cm]{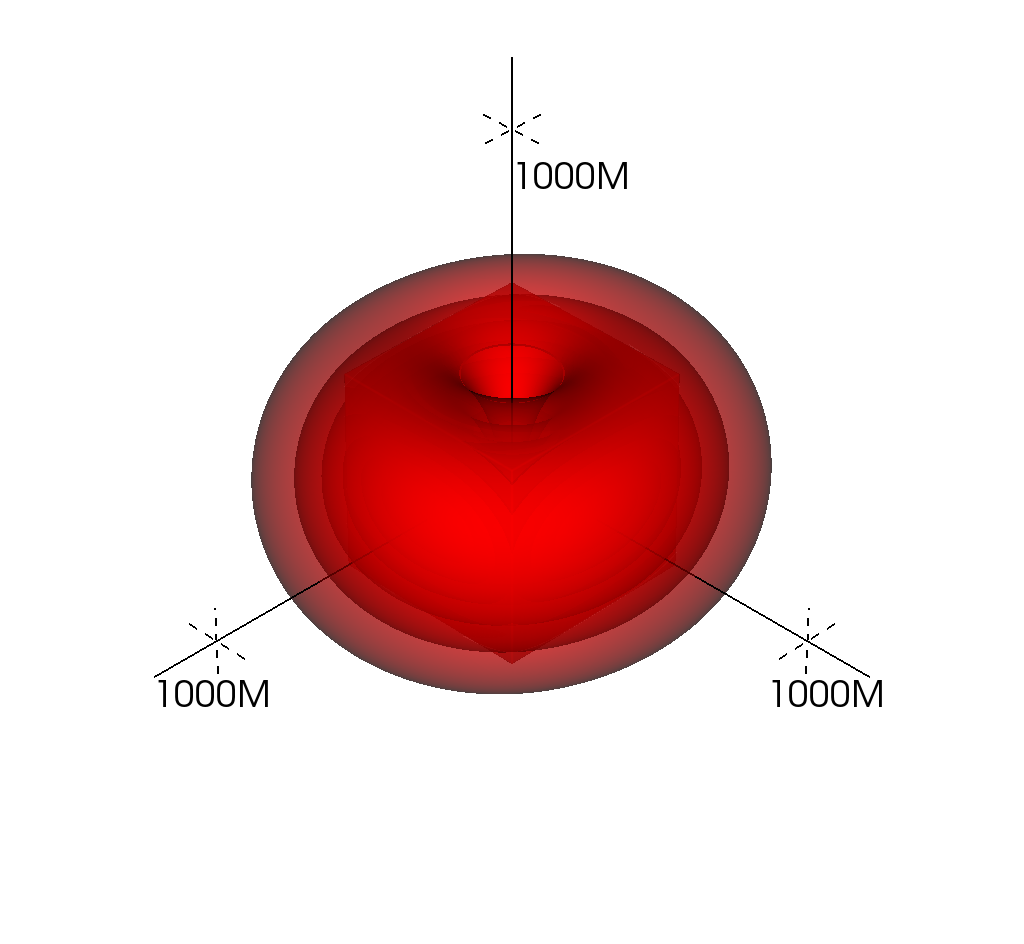}
\end{tabular}
\caption{Field (left) and energy density (right) distribution along the equatorial plane for the same initial data as Fig.~\ref{fig:a0_Mc0_mu01}. The field is dipolar, as expected, whereas the energy density at the equator is almost -- but not exactly -- symmetric along the rotation axis. The lengthscale of these images is of order $100M$.
\label{fig:a0_Mc0_mu01_snapshot}
}
\end{figure}
%

%%%%%%%%%%%%%%%%%
\section{Results}
%%%%%%%%%%%%%%%%%

%%%%%%%%%%%%%%%%%%%%%%%%%%%%%%%%%%%%%%%%%%%%%%%%%%%%%%%%%%%%%%%%%%%%%%%%%%%%
\subsection{Weak tides: transitions to new stationary states}
%%%%%%%%%%%%%%%%%%%%%%%%%%%%%%%%%%%%%%%%%%%%%%%%%%%%%%%%%%%%%%%%%%%%%%%%%%%%
%
\begin{figure}[htb]
\begin{tabular}{c}
\includegraphics[width=8.5cm]{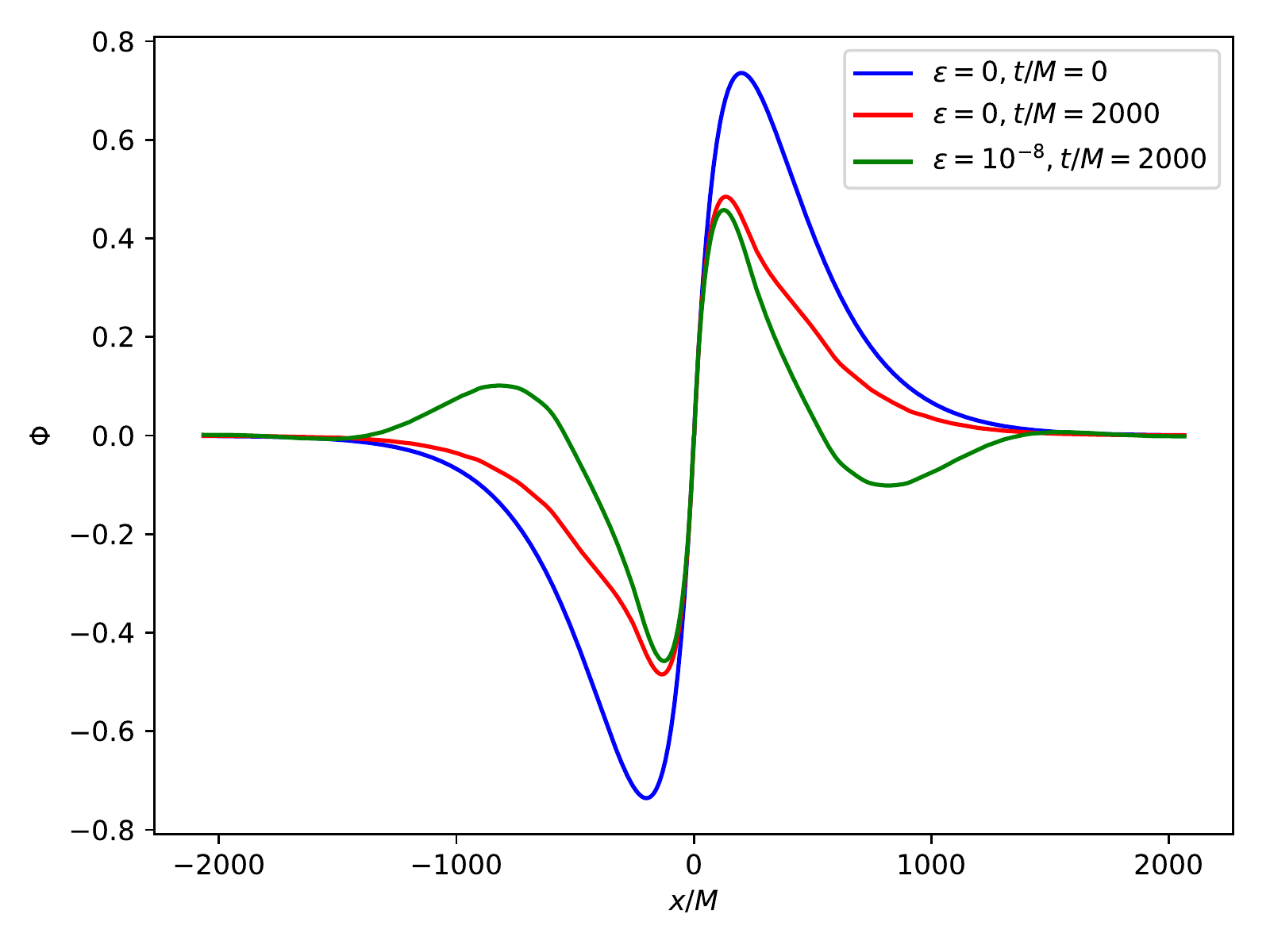}
\end{tabular}
\caption{Dependence of the field $\Phi$ along the $x-$ axis at different instants for a coupling $M\mu=0.1$. In the absence of a companion ($\epsilon=0$) and despite a slight change in the profile, the field has no nodes. It has a local extremum at $\sim r=200M$
as predicted by a small $M\mu$ expansion for the fundamental mode (in the convention adopted in this paper $n=l+1=2$).
In the presence of a weak tidal field ($\epsilon=10^{-8}$) the field develops a different radial profile with one node, pointing to a significant component of overtones. Our results indicate a sizeable excitation of the second excited state $n=4$, which has extrema at $r/M\sim 170, 850$.
\label{fig:overtones}}
\end{figure}
We start by evolving the initial data described above around an isolated, non-spinning BH ($\epsilon=0$).
The non-vanishing multipolar component of the field is shown in Fig.~\ref{fig:a0_Mc0_mu01}. The amplitude of the field varies by a few percent over the time interval of $\sim 7000M$. This time interval (7000 dynamical timescales) also corresponds to $\sim 100$
scalar field periods of oscillation. The scalar field and energy density along an equatorial slice are shown in Fig.~\ref{fig:a0_Mc0_mu01_snapshot} at $t=7000M$. The density is almost (but not exactly) symmetric along this slice.

We now turn on a weak tidal field by letting $\epsilon=10^{-8}$, produced by a companion star on the $x$-axis.
We call this a weak tide since no nonperturbative feature is seen {\it on timescales of $\sim 6000M$}. As we will argue below, it is possible that new features appear at very late times, which we are unable to probe currently.
Previous, analytical studies focused on transitions between the overtones~\cite{Baumann:2018vus,Berti:2019wnn}. We do see transitions between overtones with the same angular index, leading to an expansion of the cloud: overtones are localized at $r_{\rm Bohr}\sim n^2/(M\mu^2)$. The appearance of higher overtones is apparent in Fig.~\ref{fig:overtones}, showing the $x-$ dependence of the field initially and at $t=2000M$. 
It is clear than the companion triggers excitation of overtones, which manifest themselves via nodes in the field. 
Although this profile also includes the octupolar $l=3$ component, it is two orders of magnitude smaller than the dipolar term, as we discuss below, and unable to explain all the structure in Fig.~\ref{fig:overtones}.

The data in Figure~\ref{fig:overtones} indicates that on timescales $\lesssim 2000M$ the second excited state $n=4$ (in our convention,
states are labeled by an integer $n=l+1,l+2,...$) dominates the transitions. In fact, a small coupling $M\mu$ expansion (see Appendix~~\ref{sec:PT})
shows that the $n=3$ state has extrema at $r/M=175, 1024$ which are not apparent in Fig.~\ref{fig:overtones}.
The second excited state $n=4$ however, is predicted to have extrema at $r/M=170, 875, 2155$ in agreement with our numerical results
(but note that the last point is challenging to confirm numerically, as the grid size and spurious reflections affect a proper evaluation of eigenfunctions at large distances).

\begin{figure*}[htb]
\begin{tabular}{cc}
\includegraphics[width=8.5cm]{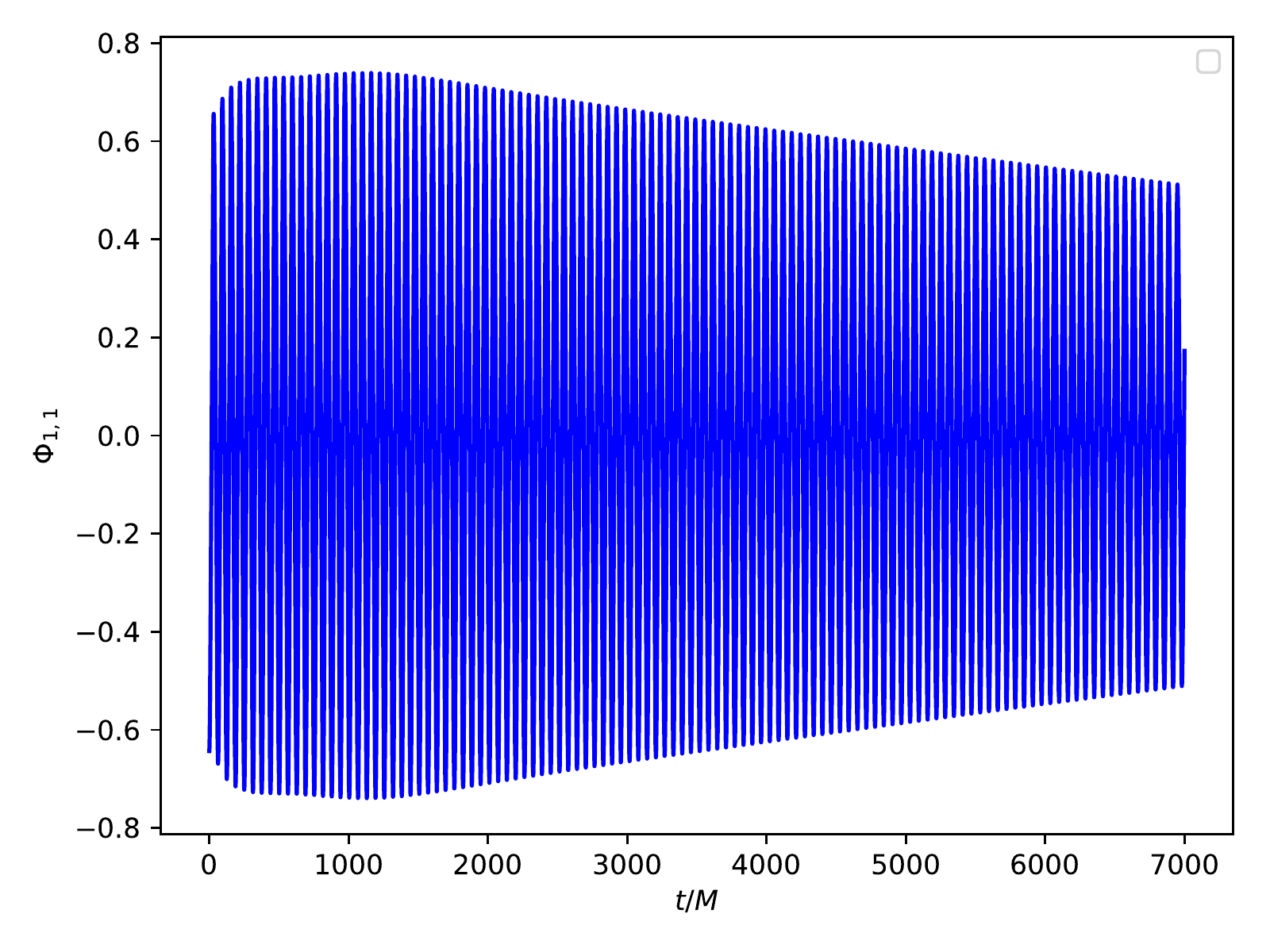}&\includegraphics[width=8.5cm]{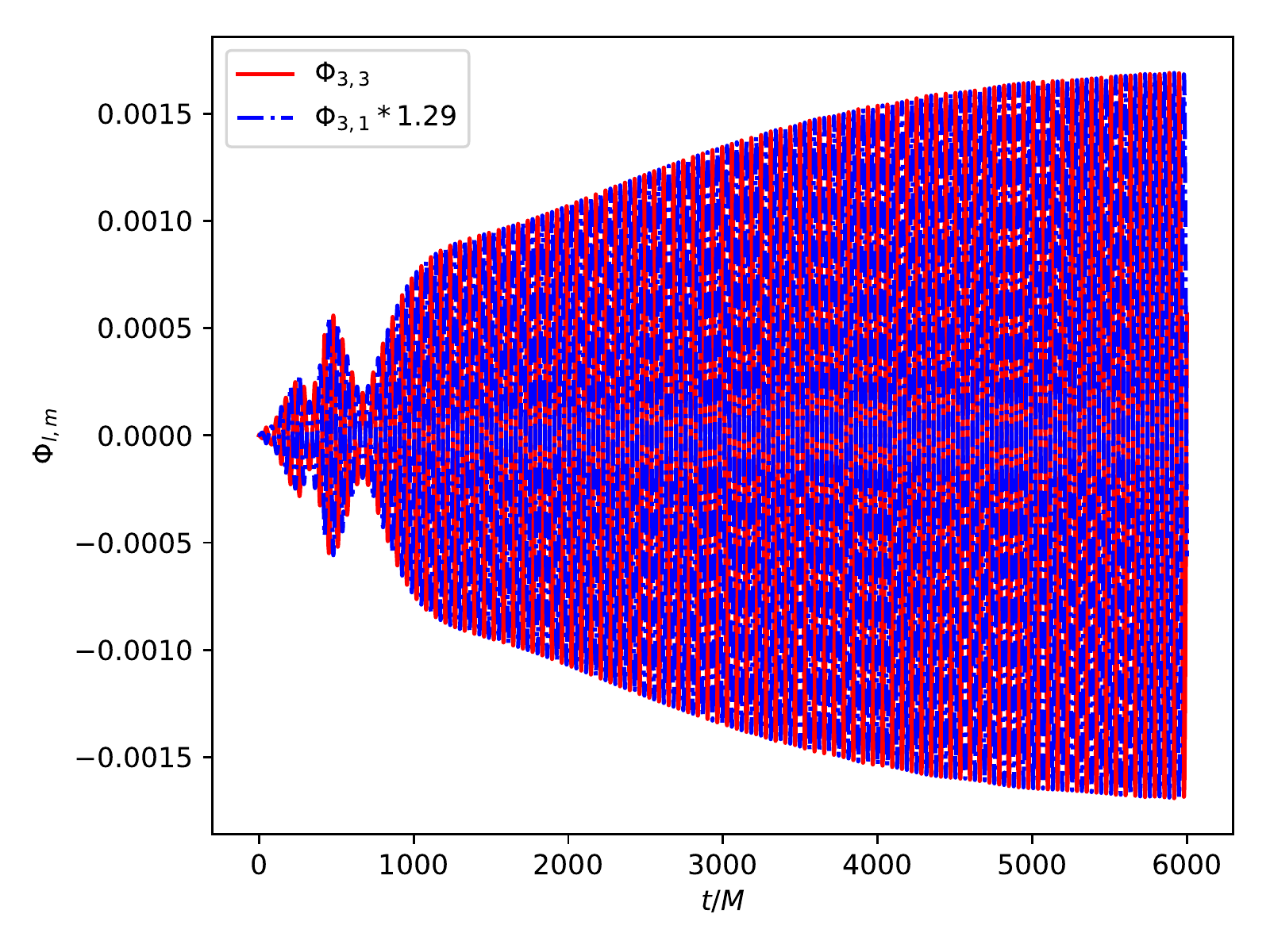}
\end{tabular}
\caption{Dipolar ($l=m=1$, left) and octupolar ($l=3,\, m=1,3$ right) component of the scalar cloud when in the presence of a weak tidal field, for the same initial conditions as in Fig.~\ref{fig:a0_Mc0_mu01} (non-spinning BH and gravitational coupling $M\mu=0.1$), but now in the presence of a companion of mass $\epsilon=10^{-8}$. The $l=3, m=1$ mode amplitude relative to the $l=m=3$ was re-scaled by the perturbation theory prediction ($\sqrt{5/3}\sim 1.29$). The agreement is very good throughout the evolution.
\label{fig:a0_Mc001_mu01}}
\end{figure*}
\begin{figure}[htb]
\begin{tabular}{c}
\includegraphics[width=8.5cm]{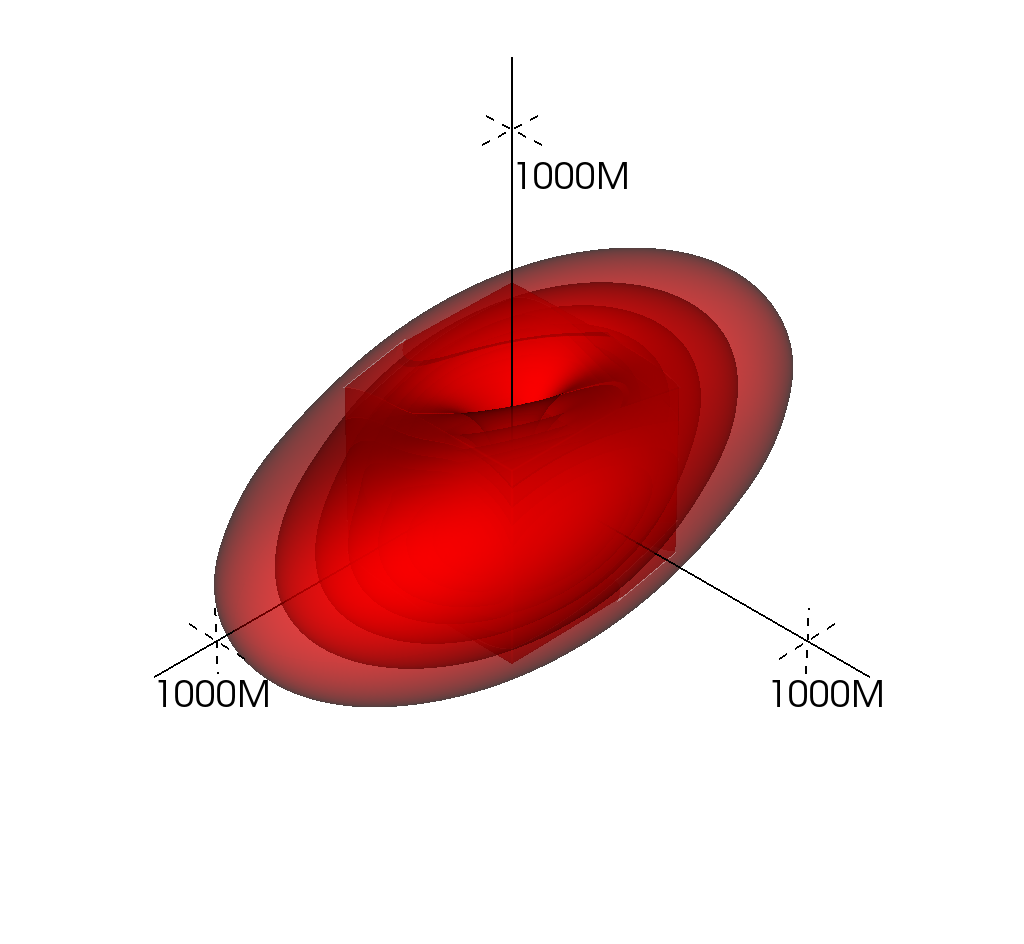}
\end{tabular}
\caption{Snapshot of a tidally deformed scalar cloud. The snapshot depicts the energy density along the equator of a scalar cloud which was set initially around a non-spinning BH. In the absence of a companion mass, the energy density is almost spherical and remains so for thousands of dynamical timescales. Here, the simulation starts with one symmetric initial scalar energy distribution, but in the presence of a star, such that the tidal parameter $\epsilon=10^{-8}$. The gravitational coupling $M\mu=0.1$.
The snapshot is taken after $7000M$ by which the system settled to a new stationary configuration. 
\label{fig:a0_Mc001_mu01_snapshot}}
\end{figure}

One can quantify the relative excitation using the orthonormality between eigenstates corresponding to different overtones, which enable us to extract the amplitude $c_n$ of a specific overtone from the numerical data via
\beq
c_n=\int_0^\infty dr \, r^2 \, R_{n1}^*\left(r\right)\Phi\left(r\right) \, , 
\label{eq:AmpOvertone}
\eeq	
where $R_{nl}\left(r\right)$ are the radial ``hydrogenic'' functions discussed in the Appendix~\ref{sec:PT} and $\Phi\left(r\right)$ corresponds to the numerical data at a given radial direction (e.g. $\theta=\pi/2$ and $\varphi=0$). We are implicitly taking the numerical data to be only composed by $l=1$ modes, which is a reasonable assumption considering our previous discussion. This expression is actually only valid in the non-relativistic (and far-region) limit, but we expect it to provide reasonable estimates at small $M\mu$ when the cloud is localized far away from the BH. 
\begin{table}[h]
\centering
\caption{Timescales $t_{\rm trans}$ and relative amplitudes predicted by time-independent perturbation theory and those obtained from numerical data (at $t=1000M$), for the most relevant $1^{st}$ order transitions from the initial state $l=1$ state, with $M\mu=0.1$. 
The second column shows the timescale to transition from the initial to the $(nlm)$ state, as obtained from time-dependent perturbation theory. The third column shows the relative amplitude of overtones, relative to the fundamental mode, from time-independent perturbation theory (and in parenthesis the corresponding ratio of the field components at $r=60M$). Finally, the last column shows the relative amplitude of overtones as obtained from our numerical data. The entries in the third and fourth column agree to within a factor two, with the exception of the $l=m=1, n=3$ mode, for which the timescale needed for excitation is larger than the instant at which the coefficients were extracted.} 
\label{tab:overtone_excitation}
\begin{tabular}{c|c|c|c}
    ($n \, l \, m$)   & $t_{\rm trans}/M$    & $\frac{c_{nlm}}{c_{211}}$  ($\frac{\phi_{nlm}}{\phi_{211}}$) &  $\frac{c_n^{\rm Num}}{c_2^{\rm Num}}$ \\
    \hline
    3 1 1   	   	   &     1888     	      & 	1.03	(0.85)	&	0.221\\
    4 1 1   	   	   &     458	          & 	0.236	(0.13)	&	0.094\\
    5 1 1   	   	   &     173	          & 	0.113	(0.046)	&	0.058\\
%    4 3 3   	   	   &     1342		      & 	0.558	($3.6\times 10^{-4}$) &	-		\\		
%    5 3 3   	   	   &     361		      & 			$1.8\times 10^{-4}$			\\		
%    6 3 3   	   	   &     148		      & 			$1.0\times 10^{-4}$			\\		
\end{tabular}
\end{table}
As we explained, our grid size is limited and for these parameters ($M\mu=0.1$), it captures a couple of nodes but not more. Thus, the relative amplitude excitation determined in this way is affected by some numerical error, and is expected to be more accurate
at early times when the signal is dominated by the fundamental mode. To avoid reflections from the outer boundary of our numerical grid and transition radius of our metric, we use time-domain data for $t\lesssim 1000M$.

Our results are shown in Table~\ref{tab:overtone_excitation}, and include estimates for the transition timescale from time-dependent perturbation theory. Our numerical results are within a factor two from the prediction from perturbation theory.
This disagreement can be explained by (at least) two factors: i. perturbation theory uses a small $M\mu$ expansion which is inaccurate at $M\mu\gtrsim 0.1$ (a glance at Fig. 1 in Ref.~\cite{Berti:2019wnn} shows how factors of two can easily arise from such an approximation); ii. transitions between intermediate states complicate substantially the calculation of mode excitation.
Bearing this in mind, our results along with perturbation theory explain why the first excited state is not yet dominant: the timescale for its excitation is the largest among those in the table. In fact, our results are consistent with transition occurring on the timescales predicted from the table for the $n=3, 4, 5$ modes.

The most apparent feature of our simulations, however, are transitions to octupolar and higher poles, induced by the external tide. This is depicted in Fig.~\ref{fig:a0_Mc001_mu01}, where we show the evolution of the dipolar $l=m=1$ and octupolar $l=m=3$ mode as time progresses. It is apparent that the magnitude of the dipolar mode is now decreasing, and that a fraction of this energy is going into higher modes, specifically the octupolar $l=3, m=1,3$. Such migration changes the spatial distribution of energy density, apparent in Fig.~\ref{fig:a0_Mc001_mu01_snapshot}.

Our results in the small external tide regime are consistent with perturbation theory prediction, in particular that the amplitude of the $l=3$ mode scales with the external tide $\epsilon$~\cite{Baumann:2018vus,Berti:2019wnn}. Perhaps one of the cleanest indications of the validity of the perturbative framework is the excitation of the $l=3,m=1$ mode. Perturbation theory predicts that the relative amplitude of the $l=m=3$ mode is $\sqrt{5/3}\sim 1.29$ larger than that of the $l=3, m=1$ mode, and this depends exclusively on an angular matrix (no radial dependence). Our numerical results show a relative amplitude across all times and extraction radii consistent with such prediction, as shown in the figure.
The $l=3$ mode seems to saturate, but as can be seen from the figure, the $l=1$ is still decreasing. Some energy is most likely flowing down the horizon, but we do not fully understand this behavior.

%%%%%%%%%%%%%%%%%%%%%%%%%%%%%%%%%%%%%%%%%%%%%%%%%%%%%%%%%%%%%%%%%%%%%%%%%%%%
\subsection{Strong tides: tidal disruption of clouds}
%%%%%%%%%%%%%%%%%%%%%%%%%%%%%%%%%%%%%%%%%%%%%%%%%%%%%%%%%%%%%%%%%%%%%%%%%%%%
\begin{figure*}[htb]
\begin{tabular}{cc}
\includegraphics[width=8.5cm]{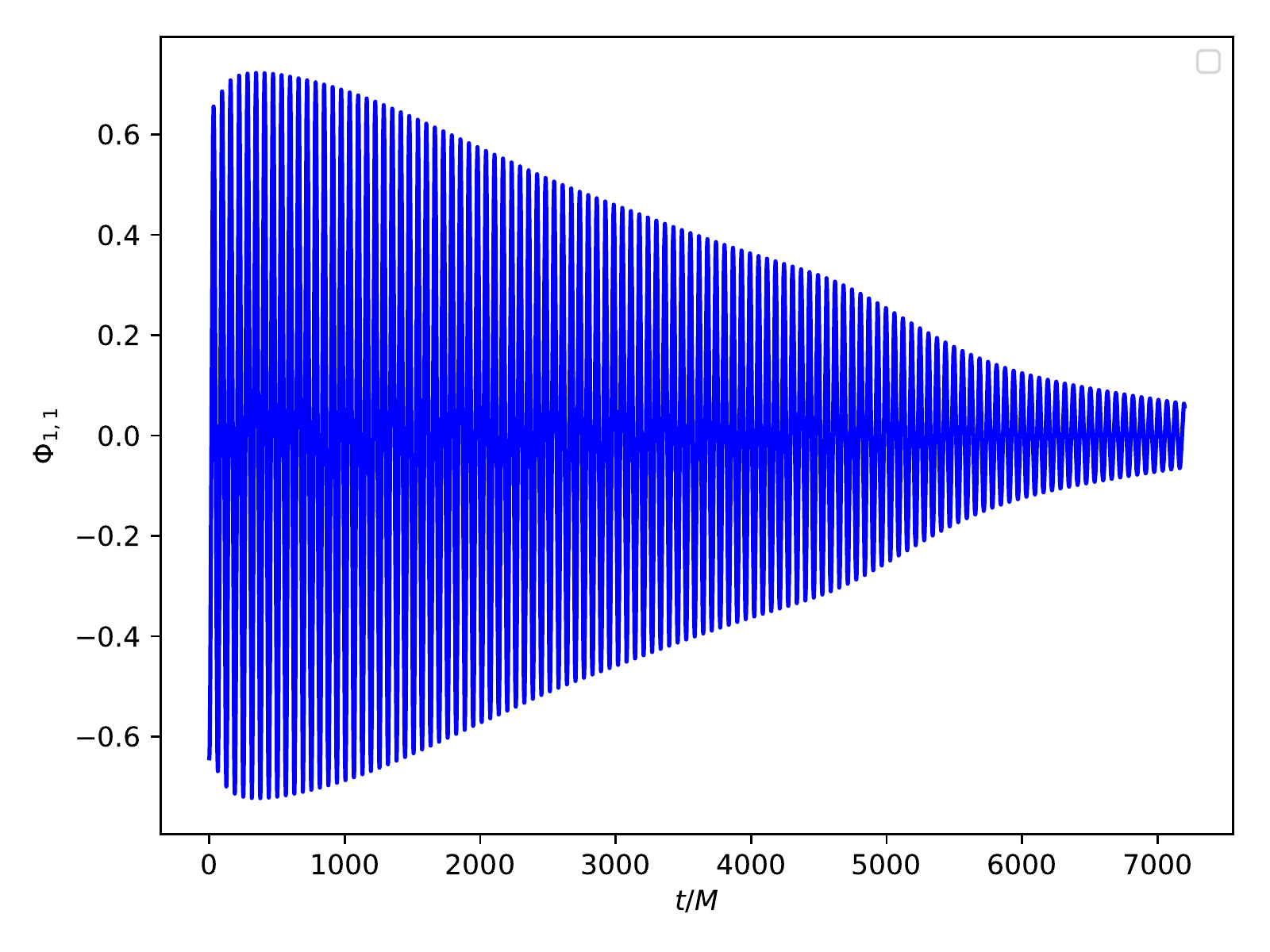}&\includegraphics[width=8.5cm]{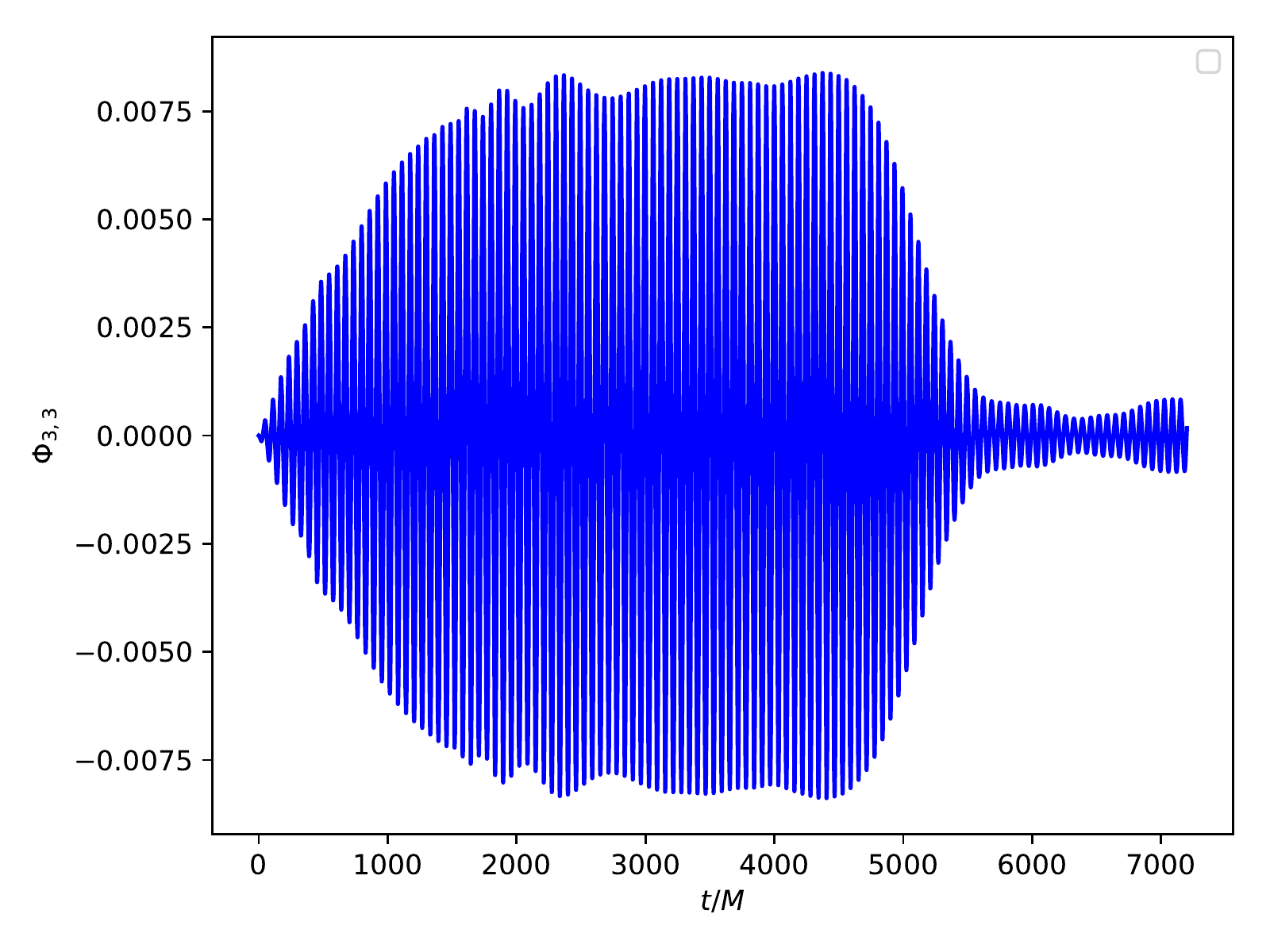}
\end{tabular}
\caption{Tidally disrupting cloud, and cascading to lower scales. This figure shows the time evolution of the dipolar and octupolar  components of the scalar field (with gravitational coupling $M\mu=0.1$) for a companion with $\epsilon=10^{-7}$ such that the cloud is disrupted. We observe that the cloud torn apart, loosing energy to asymptotically large distances, away from the BH.
Thus, energy is cascading to higher and higher multipoles as time progresses. The field is extracted at $r=60M$.
\label{fig:a0_Mc01_mu01}}
\end{figure*}
\begin{figure}[htb]
\begin{tabular}{c}
\includegraphics[width=8.5cm]{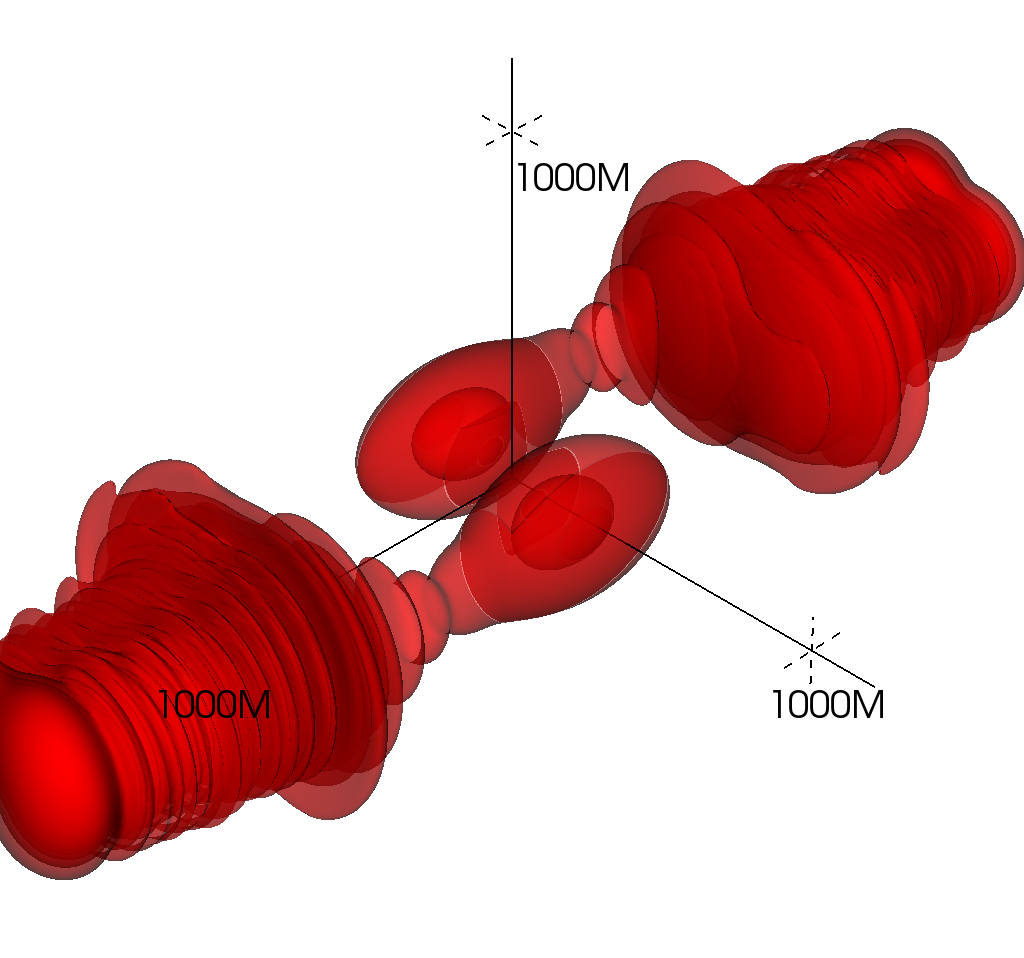}
\end{tabular}
\caption{Snapshot of a tidally disrupting cloud. The snapshot depicts the energy density along the equator of a scalar cloud which was set initially around a non-spinning BH. 
In the absence of a companion mass, the energy density is almost spherical and remains so for thousands of dynamical timescales.
Here, the simulation starts with one symmetric initial scalar energy distribution, but in the presence of a star for which $\epsilon=10^{-7}$. The gravitational coupling $M\mu=0.1$.
The snapshot is taken after $7000M$ and is leading to disruption of the cloud. 
\label{fig:a0_Mc01_mu01_snapshot}}
\end{figure}
For large tidal fields, one expects the scalar configuration to be disrupted. A star of mass $M_*$, radius $R_*$, in the presence of a companion of mass $M_c$ at distance $R$ is on the verge of disruption if -- up to numerical factors of order unity,
\be
M_*/R_*^2=2M_cR_*/R^3\,.\label{Roche}
\ee
For configurations where the mass in the scalar cloud is a fraction of that of the BH, $M_*=M$ and its radius is of the order of $R_*\gtrsim 5/(M\mu^2)$ (see Appendix). As such, we find the critical moment 
\be
\epsilon_{\rm crit}\approx\frac{(M\mu)^6}{250}\,.\label{tide_scaling}
\ee

Our simulations are consistent with this behavior. Figures~\ref{fig:a0_Mc01_mu01}--\ref{fig:a0_Mc01_mu01_snapshot}
summarize our findings at large companion masses. We find that the initial dipolar mode quickly transfers energy to the octupole,
which then drains into higher and higher multipoles. This signals a transfer of energy to lower and lower angular scales, and in our
case it is telling us that the cloud is being disrupted, loosing mass to asymptotic regions. A snapshot of the energy density in Fig.~\ref{fig:a0_Mc01_mu01_snapshot} shows precisely this.

Extracting precise values for the critical tide from our numerical simulations is complicated by two facts: these are extended scalar configurations, and to understand whether there is mass being lost to large distances requires large numerical grids. In addition, a seemingly stable cloud on some timescale can eventually be disrupted when evolved on longer timescales. Typically, our numerical simulations last for $\sim 6000M$. With this in mind, we estimate a threshold $\epsilon_{\rm crit}\sim  2\times 10^{-7}$ for $M\mu=0.2$,
for which disruption is clearly seen after $4000M$. This critical value agrees remarkably well with Eq.~\eqref{tide_scaling}.
On the other hand, for $M\mu=0.1$ we see disruption on the simulated timescales only for $\epsilon\gtrsim 2\times 10^{-8}$, a factor four difference from the prediction of Eq.~\eqref{tide_scaling}. It is possible that disruption does happen for smaller tides, but on timescales that we are currently unable to probe. Note also that disruption can be stimulated by transitions to overtones, an intermediate process which occurs as we've just discussed, and which ``puffs up'' the cloud, increasing its size to a few times the estimate $1/(M\mu^2)$ and therefore reducing the critical tide.
However, such transitions can occur on large timescales. To summarize, our results are consistent with the behavior of Eq.~\eqref{tide_scaling}, though clearly evolutions lasting for one order of magnitude more could zoom in better on the prefactor.
%%%%%%%%%%%%%%%%%%%%%%%%%%%%%%%%%%%%%%%%%%%%%%%%%%%%%%%%%%%%%%%%%%%%%%%%%%%%
\section{Application to astrophysical systems}
%%%%%%%%%%%%%%%%%%%%%%%%%%%%%%%%%%%%%%%%%%%%%%%%%%%%%%%%%%%%%%%%%%%%%%%%%%%%
Known BHs with companions include the Cygnus X-1 system and the center of our galaxy.
Cygnus X-1 is a binary system composed of a BH of mass $M_{\rm BH}\sim 15 M_\odot$, a companion with $M_c\sim 20 M_\odot$ at a distance $R\sim 0.2 \,{\rm AU} \sim 3\times 10^{10}  \,{\rm m}$~\cite{Orosz_2011}. With these parameters,
we find $\epsilon \sim 5\times 10^{-19}$. For it to sit at the critical tide, $M\mu\sim 2\times 10^{-3}$. The timescale $\tau$ for growth 
of clouds via superradiance is of order $\tau \sim (M\mu)^{-9} M$~\cite{Brito:2015oca}, too large to be meaningful for scalar fields, but potentially affecting vectors ($\tau \sim (M\mu)^{-7} M$)~\cite{Pani:2012vp,Witek:2012tr,Cardoso:2018tly}. The tide is also small enough that it should not be affecting any of the constraints derived from the possible non-observation of GWs from the system~\cite{Yoshino:2014wwa,Sun:2019mqb}.

On the other hand, at the center of our galaxy there is a supermassive BH of mass $\sim 4\times 10^6 M_{\odot}$ with known companions~\cite{Abuter:2018drb,Naoz:2019sjx}. For the closest known star, S2, with a pericenter distance of $\sim 1400M_{\rm BH}$ we find $\epsilon\sim 2\times 10^{-15}$, or a critical coupling $M\mu \sim 9\times 10^{-3}$ (we assume $M_c\sim 20 M_{\odot}$ but the result above is only mildly dependent on the unknown mass of S2). This is now a potential source of tidal disruption for interesting coupling parameters, and will certainly affect the estimates using pure dipolar modes to estimate GW emission. However, note that our approximations always require that the companion sits outside the cloud ($R>R_*$, or the approximation in Eq.~\eqref{eq:MetricPerturbation} would break down). Using Eq.~\eqref{Roche}, disruption together with such a condition always requires $M_c>M/2$.

Note that, at the verge of tidal disruption by a companion, the binary itself is emitting GWs at a rate
\be
\dot{E}_{\rm binary}= \frac{32}{5}\frac{M_c^2M^3}{R^5}\,,
\ee
where we assume the companion to be much lighter than the BH. The GW flux emitted by the cloud-BH system scales as~\cite{Yoshino:2013ofa,Brito:2014wla,Brito:2017zvb}
\beq
\dot{E}_{\text{cloud}}\sim \frac{1}{50}\left(\frac{M_S}{M}\right)^2\left(M\mu\right)^{14} \,.
\eeq
Thus, GW emission by the binary dominates the signal whenever
\be
\frac{M_c}{M_S}\gtrsim (M_S/M)^5(5M\mu/2)^{12}\,,
\ee
with $M_S$ the mass in the scalar cloud~\cite{Yoshino:2013ofa,Brito:2014wla,Brito:2017zvb}. Therefore, in the context of GW emission and detection, for all practical purposes, disruption will not affect our ability to probe the system:
if it was visible via monochromatic emission by the cloud before disruption, it will be seen after disruption as a binary.

Note that tidal disruption of the cloud is a relevant possibility for these systems, since
the cloud is generically not depleted due to mode mixing by the time the system reaches the Roche radius;
In fact, for cloud depletion due to mode mixing to be effective, the system needs to be in a resonant epoch for a long time~\cite{Baumann:2018vus}. This requires a particular combination of the mass ratio and gravitational coupling $M\mu$ which can only be realized in a small region in the possible parameter space (see Figs.~7 and 8 in Ref.~\cite{Baumann:2018vus}).

%%%%%%%%%%%%%%%%%%%%%%%%%%%%%%
\section{Conclusions}
%%%%%%%%%%%%%%%%%%%%%%%%%%%%%%
Massive, spinning BHs provide us with the tantalizing possibility to test fundamental fields on scales which are otherwise inacessible. These fields can be all or a fraction of DM and may or not couple to standard model fields.
In other words, BHs are ideal detectors of ultralight fields~\cite{Arvanitaki:2010sy,Brito:2015oca,Baumann:2019ztm}. 

The mechanism behind this extraordinary ability is superradiance, which works very much like tidal acceleration in the Earth-moon system~\cite{Cardoso:2012zn,Brito:2015oca}. In the presence of a light field, a spinning BH may transfer a large fraction of its rotational energy to a ``cloud'' of bosons orbiting the BH. Such effect leads to BH spindown, emission of nearly monochromatic radiation, etc.
We started here a numerical study of the impact of a possible companion star or BH on the development of such superradiant cloud.
We see transitions to higher overtones and to higher multipoles, {\it stretching} and {\it deforming} the cloud.
Weak tidal fields (i.e., light or far-away companions) slightly deform the cloud, affecting GW emission by the system. The changes induced by tidal fields have not been computed yet.
For tidal fields larger than the threshold of Eq.~\eqref{tide_scaling}, the companion simply breaks the cloud apart.
Since such structures are usually much larger than the BH -- as shown by Eq.~\eqref{cloud_radius} -- superradiant clouds are typically easier to disrupt than stars. In fact, BH systems such as the one at the center of our galaxy or the Cygnus X-1 binary system may easily disrupt scalar clouds.

Our results generalize to a number of situations. Although we have discussed only tides acting along the equator, we have performed evolutions for polar tides (along the $z-$axis), and found the same phenomenology. This includes overtone and transitions between multipoles and tidal disruption, even if quantitatively different.
Our setup is that of a real scalar field, but the results generalize to complex scalars. These are interesting from a BH uniqueness perspective since they can lead to truly stationary hairy solutions (as opposed to real fields which lead to long-lived states which are not of the Kerr family, but which eventually must decay to Kerr)~\cite{Herdeiro:2014goa,Brito:2015oca}.

Simulations of binaries are challenging. To overcome issues with long-term simulations of two bodies, we replace the companion with its lowest order tidal moment. Such approximation has problems of its own, and requires careful handling of boundary conditions, grid sizes etc. Currently, we are unable to probe tidal fields which vary on short timescales: these would lead to ``superluminal'' motion
on the outerpart of our computational domain; as such, we are unable to probe resonances arising from tidal effects. Resonances are interesting on their own~\cite{Baumann:2018vus,Baumann:2019ztm} and might lead to floating or sinking orbits which lead to clear imprints in GW signals~~\cite{Cardoso:2011xi,Ferreira:2017pth,Zhang:2018kib,Baumann:2019ztm}.

%%%%%%%%%%%%%%%%%%%%%%%%%%%%%%%%%%%%%%%%%%%%%%%%%%%%%%%%%%%%%%%%%%%%%%%%%%%%%
\section*{Acknowledgements}
%%%%%%%%%%%%%%%%%%%%%%%%%%%%%%%%%%%%%%%%%%%%%%%%%%%%%%%%%%%%%%%%%%%%%%%%%%%%%
%
We are grateful to Hirotaka Yoshino for useful comments and suggestions, and to
Miguel Zilh\~{a}o for useful suggestions and advice on the numerical simulations.
V. C. was partially funded by the Van der Waals Professorial Chair. V. C. would like to thank Waseda University for warm hospitality and support while this work was finalized. 
V.C. acknowledges financial support provided under the European Union's H2020 ERC 
Consolidator Grant ``Matter and strong-field gravity: New frontiers in Einstein's 
theory'' grant agreement no. MaGRaTh--646597. 
F.D. acknowledges financial support provided by FCT/Portugal through grant SFRH/BD/143657/2019.
This project has received funding from the European Union's Horizon 2020 research and innovation 
programme under the Marie Sklodowska-Curie grant agreement No 690904.
We thank FCT for financial support through Project~No.~UIDB/00099/2020.
We acknowledge financial support provided by FCT/Portugal through grant PTDC/MAT-APL/30043/2017.
The authors would like to acknowledge networking support by the GWverse COST Action 
CA16104, ``Black holes, gravitational waves and fundamental physics.''
%
%%%%%%%%%%%%%%%%%%%%%%%%%%%%%%%%%%%%%%%%%%%%%%%%%%%%%%%%%%%%%%%%%%%%%%%%%%%%%

%%%%%%%%%%%%%%%%%%%%%%%%%%%
\appendix
%%%%%%%%%%%%%%%%%%%%%%%%%%%

%%%%%%%%%%%%%%%%%%%%%%%%%%%%%%%%%%%%%%%%%%%%%%%%%%%%%%
\section{Tides in General Relativity} \label{sec:TidesGR}
%%%%%%%%%%%%%%%%%%%%%%%%%%%%%%%%%%%%%%%%%%%%%%%%%%%%%%

The general theory of tidally deformed  compact objects in General Relativity lays its foundations on linear perturbations around a background spacetime describing the compact object \cite{Binnington:2009bb, Damour:2009vw, Cardoso:2017cfl}
\beq
g_{\mu\nu}=g_{\mu\nu}^{\left(0\right)} + h_{\mu\nu} \, , \label{eq:FullMetric}
\eeq
where $g_{\mu\nu}^{\left(0\right)}$ is the background spacetime metric and $h_{\mu\nu}$ is a small perturbation. For a body perturbed by an external field, we expect $h_{\mu\nu}$ to encapsule the direct contribution of that external field and the corresponding linear response of the perturbed object due to gravitational interaction.

The standard strategy to compute $h_{\mu\nu}$ is to pick a specific gauge and solve the linearized field equations for a chosen background. The most practical situation is when the background is  spherically-symmetric and static, in which case the line element reads
\be
ds^2= -F\left(r\right)dt^2 + G\left(r\right) dr^2 + r^2 d\theta^2 + r^2 \sin^2\theta d\varphi^2 \label{eq:sphericalmetric}\, .
\ee

In this scenario, the perturbation $h_{\mu\nu}$ is expanded in spherical harmonics 
\beq
Y^{lm}(\theta,\varphi)\equiv\sqrt{\frac{2l+1}{4\pi}\frac{\left(l-m\right)!}{\left(l+m\right)!}}P^{m}_l\left(\cos\theta\right)e^{im\varphi}\, .
\eeq
and due to axisymmetry, decomposed in even and odd parts. In the Regge-Wheeler gauge, these read
\begin{widetext}
\beq
h_{\mu\nu}^{\text{even}}&=&
\begin{pmatrix}
F\left(r\right)\, H_0^{lm}\left(t,r\right)\,Y^{lm}& H_1^{lm}\left(t,r\right) \,Y^{lm} 		 & 0 & 0 \\
H_1^{lm}\left(t,r\right) \,Y^{lm} 				& 	G\left(r\right)\,H_2^{lm}\left(t,r\right) \,Y^{lm}	 & 0 & 0 \\
0 & 0 									&	r^2 \,K^{lm}\left(t,r\right) \,Y^{lm}	 & 0 \\
0 & 0 									&	0							 & r^2\sin^2\theta\, K^{lm}\left(t,r\right) \,Y^{lm}	
\end{pmatrix}\, ,\\
h_{\mu\nu}^{\text{odd}}&=&
\begin{pmatrix}
0 & 0		 & h_0^{lm}\left(t,r\right)\,S_\theta^{lm}  & h_0^{lm}\left(t,r\right)\,S_\varphi^{lm} \\
0 & 0 		 & h_1^{lm}\left(t,r\right)\,S_\theta^{lm}  & h_1^{lm}\left(t,r\right)\,S_\varphi^{lm} \\
h_0^{lm}\left(t,r\right)\,S_\theta^{lm} & h_1^{lm}\left(t,r\right)\,S_\theta^{lm} & 0 & 0\\
h_0^{lm}\left(t,r\right)\,S_\varphi^{lm}   & h_1^{lm}\left(t,r\right)\,S_\varphi^{lm}	& 0 & 0
\end{pmatrix} \, ,
\label{eq:MetricPerturbations}
\eeq
\end{widetext}
where
\beq
\left(S_\theta^{lm},S_\varphi^{lm}\right)\equiv\left(-Y^{lm}_{,\varphi}/\sin\theta,\sin\theta \, Y^{lm}_{,\theta}\right) \, .
\eeq

The aforementioned separation of $h_{\mu\nu}$ into the external field and respective tidal response can be made explicit by means of an asymptotic expansion in multipole moments (check Eqs.~(B9) and (B10) of Ref.~\cite{Damour:2009vw}). The even-parity sector is controled by the polar tidal moments $\mathcal{E}_L$, where the subscript $L\equiv a_1a_2\dots a_l$ is a multi-index labeling the $\left(2l+1\right)$ components of this symmetric-trace-free tensor. One can further decomposed these in spherical-harmonics through $\mathcal{E}_L\left(t\right) x^L = r^l \sum_m \mathcal{E}_{lm} \left(t\right) Y^{lm}\left(\theta,\varphi\right)$. The same procedure applies to the axial sector, but here they are controlled by the axial tidal moments $B_L$, which follow the same decomposition, $\mathcal{B}_L\left(t\right) x^L = \sum_m r^l \mathcal{B}_{lm} \left(t\right) Y^{lm}\left(\theta,\varphi\right)$.  

For a specific tidal field, the tidal moments can be determined by performing an asymptotic matching with a post-Newtonian environment, in a domain much larger than the typical lenghtscale of the deformed compact object \cite{Taylor:2008xy, poisson_will_2014}. We are interested in a binary system where one of bodies is a BH. Centering ourselves in it, and treating the corresponding companion as a post-Newtonian monopole of mass $M_c$ at distance $R$, the lowest contribution to the tidal field is given by the $l=2$ quadrupolar moments \cite{Taylor:2008xy}
\beq
\mathcal{E}_{ab}&=&-3\frac{M_c}{R^3}n_{\left<ab\right>} +\mathcal{O}\left(c^{-2}\right)\, ,\\
\mathcal{B}_{ab}&=&-6\frac{M_c}{R^3}\left[\textbf{n}\times \textbf{v}\right]_{(a}n_{b)}	+ \mathcal{O}\left(c^{-2}\right) \,,
\label{eq:TidalMoments}
\eeq
where $\textbf{R}$ is the position of the companion in the BH frame centered, $\text{n}\equiv \textbf{R}/R$, the brackets $\left<\dots \right>$ denote symmetrization and trace removal, and $\textbf{v}$ is the relative velocity of the binary. We stress that the two terms displayed are not of the same PN order. While the first term in $\mathcal{E}_{ab}$ corresponds to the Newtonian limit and the ommited term is a 1PN correction, the lowest order term in $\mathcal{B}_{ab}$ is already considered to be a 1PN contribution, despite not being directly surpressed by $c^{-2}$ (see section II of Ref.~\cite{Taylor:2008xy}). Higher multipole moments are also subleading with respect to above quadrupoles.

Finally, a further simplification is introduced by assuming that $h_{\mu\nu}$ is independent of time. This is the so called regime of static tides, when the binary evolution happens in much larger timescales than the internal dynamics of each body, and the whole system evolves adiabatically. This is what happens in the inspiralling phase of the binary. The corresponding perturbations for a Schwarzschild background are those presented in the main text in Eq.~\eqref{eq:MetricPerturbation}, where we took the Newtonian limit of the tidal quadrupole moments expanded in spherical harmonics, as explained above. 
%\beq
%\mathcal{B}_{ab}=-\frac{6}{R^4}L_{(a}n'_{b)} \, ,
%\eeq

%where $ \textbf{L}=M r_c \textbf{n}' \times \textbf{v}_c $ is the companion angular momentum and $\textbf{n}'=\textbf{r}'_c/r_c$.
%%%%%%%%%%%%%%%%%%%%%%%%%%%%%%%%%%%%%%%%%%%%%%%%%%%%%%%%%%%%%%%%%
\section{Perturbation Theory in Quantum Mechanics} \label{sec:PT}
%%%%%%%%%%%%%%%%%%%%%%%%%%%%%%%%%%%%%%%%%%%%%%%%%%%%%%%%%%%%%%%%%

In the non-relativistic limit, the scalar cloud obeys an equation which is formally equivalent to Schr\"{o}dinger's equation with a Coulomb potential, governed by a single parameter,
\be
\alpha\equiv M\mu \, .
\ee
This can be seen by making the standard \textit{ansatz} for the dynamical evolution of $\Phi$~\cite{Page:2003rd, Mendes:2016vdr,Baumann:2018vus} 
\beq
\Phi\left(t,\bf{r}\right)=\frac{1}{\sqrt{2\mu}}\left(\psi\left(t,\bf{r}\right)e^{-i\mu t}+\psi^*\left(t,\bf{r}\right)e^{i\mu t}\right)\,,
\eeq
where $\psi$ is a complex field which varies on timescales much larger than $1/\mu$. Then, one can re-write Eq.~\eqref{eq:MFEoMScalar} as
\beq
i \frac{\partial}{\partial t}\psi&=&\left(-\frac{1}{2\mu}\nabla^2-\frac{\alpha}{r}\right)\psi \, ,
\eeq
where we kept only terms of order $\mathcal{O}\left(r^{-1}\right)$ and linear in $\alpha$.

The normalized eigenstates of the system are hydrogenic-like, with an adapted ``fine structure constant'' $\alpha$ and ``reduced Bohr radius'' $a_0$~\cite{Detweiler:1980uk,Brito:2015oca},
\beq
\psi_{nlm}&=&\,e^{-i\left(\omega_{nlm}-\mu\right)t}\,R_{nl}\left(r\right)Y_{lm}\left(\theta,\phi\right) \, ,\\
R_{nl}\left(r\right)&=&C\left(\frac{2r}{n a_0}\right)^lL_{n-l-1}^{2l+1}\left(\frac{2r}{na_0}\right)e^{-\frac{r}{na_0}} \, , \nonumber\\
a_0&=&1/\mu\alpha \, ,\quad C=\sqrt{\left(\frac{2}{na_0}\right)^3\frac{\left(n-l-1\right)!}{2n\left(n+l\right)!}} \,,\label{eq:FrequencySpec}
\eeq
where $L_{n-l-1}^{2l+1}$ is a generalized Laguerre polynomial~\footnote{We adopt the same normalization as the built-in function of Mathematica.}. We are adopting the convention for the quantum numbers used in Refs.~\cite{Baumann:2018vus,Berti:2019wnn}.
The eigenvalue is, up to terms of order $\alpha^5$~\cite{Baumann:2019eav}
\be
\omega_{nlm}=\mu\left(1-\frac{\alpha^2}{2n^2}-\frac{\alpha^4}{8n^4}+\frac{\left(2l-3n+1\right)\alpha^4}{n^4\left(l+1/2\right)}\right)\,.
\label{eq:FrequencyEigen}
\ee
We can estimate the size of the axion cloud by computing the expectation value of the radius on a given state
\be
\left< r \right> = \int_0^{\infty}dr\,r^3 R_{nl}^2\left(r\right) = \frac{a_0}{2}\left(3n^2-l(l+1)\right) \, .\label{cloud_radius}
\ee

When the binary companion is included, the tidal perturbation can be treated in the framework of perturbation theory in Quantum Mechanics. The tidal potential $\delta V$ entering in Schr\"{o}dinger's equation due to $\delta ds^2_{\text{tidal}}$ \eqref{eq:MetricPerturbation} is represented by a step function 
\beq
\delta V=-\theta\left(t-t_0\right)\frac{M_c\, \mu}{R}\sum_{|m|\leq 2}\frac{4\pi}{5}\left(\frac{r}{R}\right)^2 Y_{lm}^*\left(\theta_c,\phi_c\right)Y_{lm}\left(\theta,\phi \right) \, , \nonumber \\
\eeq
where $t_0$ is the instant when we turn it on and $\theta\left(t\right)$ is the Heaviside function. Therefore, though there is an implicit time dependence, if one lets the system evolve for sufficient time, it will end in a final stationary state (ignoring the loss of energy at the horizon). To describe the final picture, time-independent perturbation theory is enough. 

Let us recall the standard procedure of time-independent perturbation theory. We work in the Schr\"{o}dinger's picture (though there are no ambiguities for the time-independent problem), and are trying to solve Schr\"{o}dinger's equation 
\beq
\mathcal{H}\left|\psi_i\right> &=& \omega_i\left|\psi_i \right>\, , \label{eq:SchrodingerPert}\\
\mathcal{H}&=&\mathcal{H}_0+\lambda \, \delta V \label{eq:FullHamiltonian}\, ,
\eeq
where $\mathcal{H}_0$ is the Hamiltonian of the unperturbed problem, $\delta V$ is the potential corresponding to the perturbation, and $\lambda$ is a dimensionless expansion parameter varying between $0$ (no perturbation) and $1$ (full perturbation). Since we are now referring to a generic problem, we have dropped the triple indices of the ``hydrogenic" spectrum and instead label different eigenstates $\left|\psi_i\right>$ of the Hamiltonian (and the respective eigenvalue frequencies $\omega_i$) by a single index~\footnote{In Quantum Mechanics literature, it is common to use $E$ for the (energy) eigenvalues, but since we are working in natural units, $\hbar=1$, and there is no distinction between these and the (frequency) eigenvalues being used}. 

When the system is non-degenerate, the eigenstates $\left|\psi_k^{(0)}\right>$ of the unperturbed problem (which are assumed to be known and in our case are given by Eq.~\eqref{eq:FrequencySpec}), are in one-to-one correspondence with the eingenvalues, $\omega_k^{(0)}$,

\beq
\mathcal{H}_0\left|\psi_i^{(0)}\right>=\omega_i^{(0)}\left|\psi_i^{(0)}\right> \, ,
\eeq
and $\{\psi_n^{(0)}\}$ form a complete orthonormal basis 
\beq
\left<\psi_m^{\left(0\right)}|\psi_n^{\left(0\right)}\right>=\delta_{mn}\, ,
\eeq

Now, we expand the eigenstates of the perturbed system, $\psi_i$, in terms of the basis $\{\psi_k^{\left(0\right)}\}$
\beq
\left|\psi_i\right>=\sum_k c_{ki}\left|\psi_k^{(0)}\right> \, , \label{eq:Eigenstates}
\eeq
and plugging in this \textit{ansatz} in \eqref{eq:SchrodingerPert}, the coefficients $c_{ki}$ and the eigenvalues $\omega_i$ can be obtained as a power series in $\lambda$. If the perturbation is small enough, we expect the first order expansions to be a good approximation~\cite{griffiths2017introduction}
\beq
\omega_i&=&\omega_i^{(0)}+\lambda \, \omega_i^{(1)} \, , \\
c_{ki}&=&c_{ki}^{(0)}+\lambda \, c_{ki}^{(1)} \, , \\
\omega_i^{(1)}&=&\left<\psi_i^{(0)}\right|\delta V\left|\psi_i^{(0)}\right> \, , \label{eq:EnCorr}\\
c_{ki}^{\left(1\right)}&=&\frac{\left<\psi_k^{(0)}\right|\delta V\left|\psi_i^{(0)}\right>}{\omega_i^{\left(0\right)}-\omega_k^{\left(0\right)}} \, , \, k\neq i \, , \label{eq:CoefCorr}
\eeq
where we omitted terms of order $\mathcal{O}\left(\lambda^2\right)$. In the end, we set $\lambda=1$, which is the same as reabsorbing it in $\delta V$. 

The timescales for the transitions between two modes can be estimated using time-dependent perturbation theory. This involves introducing the interaction picture and perform a Dyson series on the time-evolution operator. Since the eigenstates remain the same as in the time-independent unperturbed case, we will skip details on this procedure and directly import the result for the first-order correction on the coefficients $c_{ki}$ for a step-function perturbation \cite{sakurai2011modern}
\beq
c_{ki}^{\left(1\right)}&=&\frac{\left<\psi_k\right|\delta V\left|\psi_i\right>}{\omega_i-\omega_k}\left(1-e^{-i\left(\omega_i-\omega_k\right)t}\right) \, .
\eeq
Both the states $\left|\psi_i\right>$ and frequencies $\omega_i$ should be understood as the ones for the unperturbed system, but we ommit subscripts to avoid clustering, Then, the probability  of the transition $\left|i\right> \rightarrow \left|k\right>$ is
\beq
\left|c_{ki}^{\left(1\right)}\right|^2&=&4\left|\frac{\left<\psi_k\right|\delta V\left|\psi_i\right>}{\omega_i-\omega_k}\right|^2\sin^2\left(\frac{(\omega_k-\omega_i)t}{2}\right) \, .
\eeq

Although we do not have a continuum spectrum, for large timescales we can take this limit. Then, at fixed $t$, we can treat the probabilities $\left|c_{ki}^{\left(1\right)}\right|^2$ as functions of
\beq
\Delta \omega_{ki}=\left|\omega_k-\omega_i\right|\, .
\eeq

Plotting it for different instants of time, one can verify this function becomes increasingly peaked around $\Delta \omega_{ki}=0$ as $t$ increases (check Fig. 5.8 of Ref.~\cite{sakurai2011modern}). This central peak scales with $t^2$ and has a typical width of $1/t$. If we wait enough time $\Delta t$ since the perturbation is introduced, the only transitions with appreciable probability are those satisfying
\beq
\Delta t=2\pi/\Delta \omega_{ki} \, .
\eeq

The final conclusion is that the typical timescale $\Delta t$ for the transition $\left|i\right> \rightarrow \left|k\right>$ to happen is
\beq
\Delta \omega_{ki} \, \Delta t \, \sim \, 1 \, ,
\label{eq:Timescales}
\eeq
which, if we momentarily insert factors of $\hbar$, can be seen as a manifestation of the energy-time uncertainty principle \cite{sakurai2011modern}
\beq
\Delta E \Delta t \sim \hbar \, .
\eeq

Returning to our problem, the initial data \eqref{Eq.axion cloud initial data} corresponds to the stationary state (reintroducing the triple ``hydrogenic"" indices) 
\beq
\left|i \right> \propto \left(\left|\psi^{(0)}_{211}\right> - \left|\psi^{(0)}_{21-1}\right> \right) \, , 
\label{eq:InitialState}
\eeq
up to a proportionality constant reflecting the renormalization done for numerical purposes. The final state should correspond to a stationary state $\left|f\right>$ which we can compute using the machinery developed before
%
%\beq
%\left|f \right> \propto \left(\left|\psi_{211}\right> - \left|\psi_{21-1}\right> \right) \, ,
%\label{eq:FinalState}
%\eeq
% 
%where these are eigenstates of \eqref{eq:FullHamiltonian}, which we can compute using the machinery developed before. 
There is still a \textit{caveat}, which is the degeneracy between states with the same quantum number $m$ \eqref{eq:FrequencySpec}. Though a rotating BH will lift this degeneracy, the energy shifts due to the perturbations considered are orders of magnitude higher than the energy scale associated with the rotation. Thus, for non-degenerate perturbation theory to be controlled, we would have to perform it at higher orders then what we presented. 

In the degenerate scenario, the equations presented are invalid (for example \eqref{eq:CoefCorr} diverges when $\omega_k^{\left(0\right)}=\omega_i^{\left(0\right)}$). Instead, we use the freedom in making a linear combination of unperturbed degenerate eigenstates, so that in every degenerate subspace, we pick a basis of the Hilbert space that diagonalizes the full Hamiltonian $\mathcal{H}$ \eqref{eq:FullHamiltonian}. After this step, we can apply non-degenerate perturbation theory, namely Eqs.~\eqref{eq:EnCorr} and \eqref{eq:CoefCorr}, using the new ``good'' basis.

Finally, the numerical data we present in the main text corresponds to multipole expansions of the field $\Phi$ and not to the coefficients $c_{ki}$ \eqref{eq:Eigenstates} describing the mix of the unperturbed states. To obtain these multipoles we have to select them from the space representation of the final state. The amplitude coefficients of the mode $\left|\psi_{nlm}\right>$ are obtained via
\beq
c_{nlm}&\propto& \frac{\left<\psi_{nlm}\left|\delta V\right|i\right>}{{\omega_{21}^{\left(0\right)}-\omega_{nl}^{\left(0\right)}}} \, , \\
\phi_{nlm}\left(r\right) &\propto& c_{nlm} R_{nl}\left(r\right) \, .
\eeq

In the end, we are interested in the ratio between amplitudes so the constant of proportionality is irrelevant. The matrix elements appearing here are explicitly presented in Eqs.(3.7)-(3.9) of Ref.~\cite{Baumann:2018vus}. Notice that the relative amplitude between modes with the same quantum numbers $n$ and $l$ is completely determined by the angular integrals, and since these are (quasi)degenerate, they will also follow similar time evolutions. As a consequence, their relative amplitude  is independent of time and the value of $\alpha$, even at higher orders in perturbation theory. This is illustrated in Fig.~\ref{fig:a0_Mc001_mu01} for $\phi_{n33}/\phi_{n31}$.

A summary of the time-independent perturbation theory for transitions between overtones is shown in Table~\ref{tab:overtone_excitation} for $M\mu=0.1$, $\epsilon=10^{-8}$. The relative amplitudes $c_{nlm}/c_{211}$ indicate that the perturbation is not that small. 
This is even more obvious if we compute the first order corrections to the frequency eigenvalues (B16) which, for this configuration, are of $\mathcal{O}\left(10^{-3}\right)$ for overtones $n>3$, as illustrated in Table~\ref{tab:CorrectedFreq}. For this reason, when computing the timescales of the transitions \eqref{eq:Timescales}, we used the first order corrected  $\omega_{nlm}$.

\begin{table}[h]
\centering
\caption{First order corrected frequencies $\omega_{nlm}$ predicted by time-independent theory, for a non-rotating BH and a companion with the configuration $M\mu=0.1$, $\epsilon=10^{-8}$. A spinning BH would break the degeneracy between states with the same $l$ but different $m$ quantum number. However, these corrections enter the frequency spectrum \eqref{eq:FrequencyEigen} only at order $\alpha^5$. For the above configuration, these would yield $\omega_{n33}-\omega_{n31} \sim 10^{-6}a/M\,n^3$, where $a$ is the angular momentum parameter $a=J/M$.}
\begin{tabular}{c|c}
    ($n \, l $)   & $\omega_{nlm} \times 10^2$   \\
    \hline
    2 1    	   	   &        9.9754	      	\\
    3 1    	   	   &        9.9224  	    \\
    4 1    	   	   &     	9.7570          \\
    5 1    	   	   &     	9.3980          \\
    4 3 	   	   &     	9.9001         	\\
\end{tabular}
\label{tab:CorrectedFreq}
\end{table}
\bibliographystyle{utphys}
\bibliography{References}

\providecommand{\href}[2]{#2}\begingroup\raggedright\begin{thebibliography}{10}

\bibitem{Bertone:2018xtm}
G.~Bertone and M.~P. Tait, Tim, ``{A new era in the search for dark matter},''
  \href{http://dx.doi.org/10.1038/s41586-018-0542-z}{{\em Nature} {\bfseries
  562} no.~7725, (2018) 51--56},
\href{http://arxiv.org/abs/1810.01668}{{\ttfamily arXiv:1810.01668
  [astro-ph.CO]}}.
%%CITATION = ARXIV:1810.01668;%%.

\bibitem{Barack:2018yly}
L.~Barack {\em et~al.}, ``{Black holes, gravitational waves and fundamental
  physics: a roadmap},'' \href{http://dx.doi.org/10.1088/1361-6382/ab0587}{{\em
  Class. Quant. Grav.} {\bfseries 36} no.~14, (2019) 143001},
\href{http://arxiv.org/abs/1806.05195}{{\ttfamily arXiv:1806.05195 [gr-qc]}}.
%%CITATION = ARXIV:1806.05195;%%.

\bibitem{Baibhav:2019rsa}
V.~Baibhav {\em et~al.}, ``{Probing the Nature of Black Holes: Deep in the mHz
  Gravitational-Wave Sky},''
\href{http://arxiv.org/abs/1908.11390}{{\ttfamily arXiv:1908.11390
  [astro-ph.HE]}}.
%%CITATION = ARXIV:1908.11390;%%.

\bibitem{Maggiore:2019uih}
M.~Maggiore {\em et~al.}, ``{Science Case for the Einstein Telescope},''
\href{http://arxiv.org/abs/1912.02622}{{\ttfamily arXiv:1912.02622
  [astro-ph.CO]}}.
%%CITATION = ARXIV:1912.02622;%%.

\bibitem{Cardoso:2016ryw}
V.~Cardoso and L.~Gualtieri, ``{Testing the black hole ‘no-hair’
  hypothesis},'' \href{http://dx.doi.org/10.1088/0264-9381/33/17/174001}{{\em
  Class. Quant. Grav.} {\bfseries 33} no.~17, (2016) 174001},
\href{http://arxiv.org/abs/1607.03133}{{\ttfamily arXiv:1607.03133 [gr-qc]}}.
%%CITATION = ARXIV:1607.03133;%%.

\bibitem{Brito:2015oca}
R.~Brito, V.~Cardoso, and P.~Pani, ``{Superradiance},''
  \href{http://dx.doi.org/10.1007/978-3-319-19000-6}{{\em Lect. Notes Phys.}
  {\bfseries 906} (2015) pp.1--237},
\href{http://arxiv.org/abs/1501.06570}{{\ttfamily arXiv:1501.06570 [gr-qc]}}.
%%CITATION = ARXIV:1501.06570;%%.

\bibitem{Eda:2014kra}
K.~Eda, Y.~Itoh, S.~Kuroyanagi, and J.~Silk, ``{Gravitational waves as a probe
  of dark matter minispikes},''
  \href{http://dx.doi.org/10.1103/PhysRevD.91.044045}{{\em Phys. Rev.}
  {\bfseries D91} no.~4, (2015) 044045},
\href{http://arxiv.org/abs/1408.3534}{{\ttfamily arXiv:1408.3534 [gr-qc]}}.
%%CITATION = ARXIV:1408.3534;%%.

\bibitem{Macedo:2013qea}
C.~F.~B. Macedo, P.~Pani, V.~Cardoso, and L.~C.~B. Crispino, ``{Into the lair:
  gravitational-wave signatures of dark matter},''
  \href{http://dx.doi.org/10.1088/0004-637X/774/1/48}{{\em Astrophys. J.}
  {\bfseries 774} (2013) 48},
\href{http://arxiv.org/abs/1302.2646}{{\ttfamily arXiv:1302.2646 [gr-qc]}}.
%%CITATION = ARXIV:1302.2646;%%.

\bibitem{Barausse:2014tra}
E.~Barausse, V.~Cardoso, and P.~Pani, ``{Can environmental effects spoil
  precision gravitational-wave astrophysics?},''
  \href{http://dx.doi.org/10.1103/PhysRevD.89.104059}{{\em Phys. Rev.}
  {\bfseries D89} no.~10, (2014) 104059},
\href{http://arxiv.org/abs/1404.7149}{{\ttfamily arXiv:1404.7149 [gr-qc]}}.
%%CITATION = ARXIV:1404.7149;%%.

\bibitem{Cardoso:2019rou}
V.~Cardoso and A.~Maselli, ``{Constraints on the astrophysical environment of
  binaries with gravitational-wave observations},''
\href{http://arxiv.org/abs/1909.05870}{{\ttfamily arXiv:1909.05870
  [astro-ph.HE]}}.
%%CITATION = ARXIV:1909.05870;%%.

\bibitem{Arvanitaki:2009fg}
A.~Arvanitaki, S.~Dimopoulos, S.~Dubovsky, N.~Kaloper, and J.~March-Russell,
  ``{String Axiverse},''
  \href{http://dx.doi.org/10.1103/PhysRevD.81.123530}{{\em Phys. Rev.}
  {\bfseries D81} (2010) 123530},
\href{http://arxiv.org/abs/0905.4720}{{\ttfamily arXiv:0905.4720 [hep-th]}}.
%%CITATION = ARXIV:0905.4720;%%.

\bibitem{Marsh:2015xka}
D.~J.~E. Marsh, ``{Axion Cosmology},''
  \href{http://dx.doi.org/10.1016/j.physrep.2016.06.005}{{\em Phys. Rept.}
  {\bfseries 643} (2016) 1--79},
\href{http://arxiv.org/abs/1510.07633}{{\ttfamily arXiv:1510.07633
  [astro-ph.CO]}}.
%%CITATION = ARXIV:1510.07633;%%.

\bibitem{Arvanitaki:2010sy}
A.~Arvanitaki and S.~Dubovsky, ``{Exploring the String Axiverse with Precision
  Black Hole Physics},''
  \href{http://dx.doi.org/10.1103/PhysRevD.83.044026}{{\em Phys. Rev.}
  {\bfseries D83} (2011) 044026},
\href{http://arxiv.org/abs/1004.3558}{{\ttfamily arXiv:1004.3558 [hep-th]}}.
%%CITATION = ARXIV:1004.3558;%%.

\bibitem{Brito:2014wla}
R.~Brito, V.~Cardoso, and P.~Pani, ``{Black holes as particle detectors:
  evolution of superradiant instabilities},''
  \href{http://dx.doi.org/10.1088/0264-9381/32/13/134001}{{\em Class. Quant.
  Grav.} {\bfseries 32} no.~13, (2015) 134001},
\href{http://arxiv.org/abs/1411.0686}{{\ttfamily arXiv:1411.0686 [gr-qc]}}.
%%CITATION = ARXIV:1411.0686;%%.

\bibitem{Arvanitaki:2016qwi}
A.~Arvanitaki, M.~Baryakhtar, S.~Dimopoulos, S.~Dubovsky, and R.~Lasenby,
  ``{Black Hole Mergers and the QCD Axion at Advanced LIGO},''
  \href{http://dx.doi.org/10.1103/PhysRevD.95.043001}{{\em Phys. Rev.}
  {\bfseries D95} no.~4, (2017) 043001},
\href{http://arxiv.org/abs/1604.03958}{{\ttfamily arXiv:1604.03958 [hep-ph]}}.
%%CITATION = ARXIV:1604.03958;%%.

\bibitem{Brito:2017zvb}
R.~Brito, S.~Ghosh, E.~Barausse, E.~Berti, V.~Cardoso, I.~Dvorkin, A.~Klein,
  and P.~Pani, ``{Gravitational wave searches for ultralight bosons with LIGO
  and LISA},'' \href{http://dx.doi.org/10.1103/PhysRevD.96.064050}{{\em Phys.
  Rev.} {\bfseries D96} no.~6, (2017) 064050},
\href{http://arxiv.org/abs/1706.06311}{{\ttfamily arXiv:1706.06311 [gr-qc]}}.
%%CITATION = ARXIV:1706.06311;%%.

\bibitem{Brito:2017wnc}
R.~Brito, S.~Ghosh, E.~Barausse, E.~Berti, V.~Cardoso, I.~Dvorkin, A.~Klein,
  and P.~Pani, ``{Stochastic and resolvable gravitational waves from ultralight
  bosons},'' \href{http://dx.doi.org/10.1103/PhysRevLett.119.131101}{{\em Phys.
  Rev. Lett.} {\bfseries 119} no.~13, (2017) 131101},
\href{http://arxiv.org/abs/1706.05097}{{\ttfamily arXiv:1706.05097 [gr-qc]}}.
%%CITATION = ARXIV:1706.05097;%%.

\bibitem{Ferreira:2017pth}
M.~C. Ferreira, C.~F.~B. Macedo, and V.~Cardoso, ``{Orbital fingerprints of
  ultralight scalar fields around black holes},''
  \href{http://dx.doi.org/10.1103/PhysRevD.96.083017}{{\em Phys. Rev.}
  {\bfseries D96} no.~8, (2017) 083017},
\href{http://arxiv.org/abs/1710.00830}{{\ttfamily arXiv:1710.00830 [gr-qc]}}.
%%CITATION = ARXIV:1710.00830;%%.

\bibitem{Boskovic:2018rub}
M.~Boskovic, F.~Duque, M.~C. Ferreira, F.~S. Miguel, and V.~Cardoso, ``{Motion
  in time-periodic backgrounds with applications to ultralight dark matter
  haloes at galactic centers},''
  \href{http://dx.doi.org/10.1103/PhysRevD.98.024037}{{\em Phys. Rev.}
  {\bfseries D98} (2018) 024037},
\href{http://arxiv.org/abs/1806.07331}{{\ttfamily arXiv:1806.07331 [gr-qc]}}.
%%CITATION = ARXIV:1806.07331;%%.

\bibitem{Cardoso:2011xi}
V.~Cardoso, S.~Chakrabarti, P.~Pani, E.~Berti, and L.~Gualtieri, ``{Floating
  and sinking: The Imprint of massive scalars around rotating black holes},''
  \href{http://dx.doi.org/10.1103/PhysRevLett.107.241101}{{\em Phys. Rev.
  Lett.} {\bfseries 107} (2011) 241101},
\href{http://arxiv.org/abs/1109.6021}{{\ttfamily arXiv:1109.6021 [gr-qc]}}.
%%CITATION = ARXIV:1109.6021;%%.

\bibitem{Zhang:2018kib}
J.~Zhang and H.~Yang, ``{Gravitational floating orbits around hairy black
  holes},'' \href{http://dx.doi.org/10.1103/PhysRevD.99.064018}{{\em Phys.
  Rev.} {\bfseries D99} no.~6, (2019) 064018},
\href{http://arxiv.org/abs/1808.02905}{{\ttfamily arXiv:1808.02905 [gr-qc]}}.
%%CITATION = ARXIV:1808.02905;%%.

\bibitem{Zhang:2019eid}
J.~Zhang and H.~Yang, ``{Dynamic Signatures of Black Hole Binaries with
  Superradiant Clouds},''
\href{http://arxiv.org/abs/1907.13582}{{\ttfamily arXiv:1907.13582 [gr-qc]}}.
%%CITATION = ARXIV:1907.13582;%%.

\bibitem{Baumann:2019ztm}
D.~Baumann, H.~S. Chia, R.~A. Porto, and J.~Stout, ``{Gravitational Collider
  Physics},''
\href{http://arxiv.org/abs/1912.04932}{{\ttfamily arXiv:1912.04932 [gr-qc]}}.
%%CITATION = ARXIV:1912.04932;%%.

\bibitem{Ikeda:2019fvj}
T.~Ikeda, R.~Brito, and V.~Cardoso, ``{Blasts of Light from Axions},''
  \href{http://dx.doi.org/10.1103/PhysRevLett.122.081101}{{\em Phys. Rev.
  Lett.} {\bfseries 122} no.~8, (2019) 081101},
\href{http://arxiv.org/abs/1811.04950}{{\ttfamily arXiv:1811.04950 [gr-qc]}}.
%%CITATION = ARXIV:1811.04950;%%.

\bibitem{Boskovic:2018lkj}
M.~Boskovic, R.~Brito, V.~Cardoso, T.~Ikeda, and H.~Witek, ``{Axionic
  instabilities and new black hole solutions},''
  \href{http://dx.doi.org/10.1103/PhysRevD.99.035006}{{\em Phys. Rev.}
  {\bfseries D99} no.~3, (2019) 035006},
\href{http://arxiv.org/abs/1811.04945}{{\ttfamily arXiv:1811.04945 [gr-qc]}}.
%%CITATION = ARXIV:1811.04945;%%.

\bibitem{Arvanitaki:2014wva}
A.~Arvanitaki, M.~Baryakhtar, and X.~Huang, ``{Discovering the QCD Axion with
  Black Holes and Gravitational Waves},''
  \href{http://dx.doi.org/10.1103/PhysRevD.91.084011}{{\em Phys. Rev.}
  {\bfseries D91} no.~8, (2015) 084011},
\href{http://arxiv.org/abs/1411.2263}{{\ttfamily arXiv:1411.2263 [hep-ph]}}.
%%CITATION = ARXIV:1411.2263;%%.

\bibitem{Baumann:2018vus}
D.~Baumann, H.~S. Chia, and R.~A. Porto, ``{Probing Ultralight Bosons with
  Binary Black Holes},''
  \href{http://dx.doi.org/10.1103/PhysRevD.99.044001}{{\em Phys. Rev.}
  {\bfseries D99} no.~4, (2019) 044001},
\href{http://arxiv.org/abs/1804.03208}{{\ttfamily arXiv:1804.03208 [gr-qc]}}.
%%CITATION = ARXIV:1804.03208;%%.

\bibitem{Berti:2019wnn}
E.~Berti, R.~Brito, C.~F.~B. Macedo, G.~Raposo, and J.~L. Rosa, ``{Ultralight
  boson cloud depletion in binary systems},''
  \href{http://dx.doi.org/10.1103/PhysRevD.99.104039}{{\em Phys. Rev.}
  {\bfseries D99} no.~10, (2019) 104039},
\href{http://arxiv.org/abs/1904.03131}{{\ttfamily arXiv:1904.03131 [gr-qc]}}.
%%CITATION = ARXIV:1904.03131;%%.

\bibitem{Cardoso:2012zn}
V.~Cardoso and P.~Pani, ``{Tidal acceleration of black holes and
  superradiance},'' \href{http://dx.doi.org/10.1088/0264-9381/30/4/045011}{{\em
  Class. Quant. Grav.} {\bfseries 30} (2013) 045011},
\href{http://arxiv.org/abs/1205.3184}{{\ttfamily arXiv:1205.3184 [gr-qc]}}.
%%CITATION = ARXIV:1205.3184;%%.

\bibitem{Cardoso:2017cfl}
V.~Cardoso, E.~Franzin, A.~Maselli, P.~Pani, and G.~Raposo, ``{Testing
  strong-field gravity with tidal Love numbers},''
  \href{http://dx.doi.org/10.1103/PhysRevD.95.084014}{{\em Phys. Rev.}
  {\bfseries D95} no.~8, (2017) 084014},
\href{http://arxiv.org/abs/1701.01116}{{\ttfamily arXiv:1701.01116 [gr-qc]}}.
%%CITATION = ARXIV:1701.01116;%%.

\bibitem{Cardoso:2019upw}
V.~Cardoso and F.~Duque, ``{Environmental effects in GW physics: tidal
  deformability of black holes immersed in matter},''
\href{http://arxiv.org/abs/1912.07616}{{\ttfamily arXiv:1912.07616 [gr-qc]}}.
%%CITATION = ARXIV:1912.07616;%%.

\bibitem{Taylor:2008xy}
S.~Taylor and E.~Poisson, ``{Nonrotating black hole in a post-Newtonian tidal
  environment},'' \href{http://dx.doi.org/10.1103/PhysRevD.78.084016}{{\em
  Phys. Rev.} {\bfseries D78} (2008) 084016},
\href{http://arxiv.org/abs/0806.3052}{{\ttfamily arXiv:0806.3052 [gr-qc]}}.
%%CITATION = ARXIV:0806.3052;%%.

\bibitem{Witek:2012tr}
H.~Witek, V.~Cardoso, A.~Ishibashi, and U.~Sperhake, ``{Superradiant
  instabilities in astrophysical systems},''
  \href{http://dx.doi.org/10.1103/PhysRevD.87.043513}{{\em Phys. Rev.}
  {\bfseries D87} no.~4, (2013) 043513},
\href{http://arxiv.org/abs/1212.0551}{{\ttfamily arXiv:1212.0551 [gr-qc]}}.
%%CITATION = ARXIV:1212.0551;%%.

\bibitem{Yoshino:2013ofa}
H.~Yoshino and H.~Kodama, ``{Gravitational radiation from an axion cloud around
  a black hole: Superradiant phase},''
  \href{http://dx.doi.org/10.1093/ptep/ptu029}{{\em PTEP} {\bfseries 2014}
  (2014) 043E02},
\href{http://arxiv.org/abs/1312.2326}{{\ttfamily arXiv:1312.2326 [gr-qc]}}.
%%CITATION = ARXIV:1312.2326;%%.

\bibitem{Orosz_2011}
J.~A. Orosz, J.~E. McClintock, J.~P. Aufdenberg, R.~A. Remillard, M.~J. Reid,
  R.~Narayan, and L.~Gou, ``The mass of the black hole in cygnus x-1,''
  \href{http://dx.doi.org/10.1088/0004-637x/742/2/84}{{\em The Astrophysical
  Journal} {\bfseries 742} no.~2, (Nov, 2011) 84}.
  \url{http://dx.doi.org/10.1088/0004-637X/742/2/84}.

\bibitem{Pani:2012vp}
P.~Pani, V.~Cardoso, L.~Gualtieri, E.~Berti, and A.~Ishibashi, ``{Black hole
  bombs and photon mass bounds},''
  \href{http://dx.doi.org/10.1103/PhysRevLett.109.131102}{{\em Phys. Rev.
  Lett.} {\bfseries 109} (2012) 131102},
\href{http://arxiv.org/abs/1209.0465}{{\ttfamily arXiv:1209.0465 [gr-qc]}}.
%%CITATION = ARXIV:1209.0465;%%.

\bibitem{Cardoso:2018tly}
V.~Cardoso, O.~J.~C. Dias, G.~S. Hartnett, M.~Middleton, P.~Pani, and J.~E.
  Santos, ``{Constraining the mass of dark photons and axion-like particles
  through black-hole superradiance},''
  \href{http://dx.doi.org/10.1088/1475-7516/2018/03/043}{{\em JCAP} {\bfseries
  1803} no.~03, (2018) 043},
\href{http://arxiv.org/abs/1801.01420}{{\ttfamily arXiv:1801.01420 [gr-qc]}}.
%%CITATION = ARXIV:1801.01420;%%.

\bibitem{Yoshino:2014wwa}
H.~Yoshino and H.~Kodama, ``{Probing the string axiverse by gravitational waves
  from Cygnus X-1},'' \href{http://dx.doi.org/10.1093/ptep/ptv067}{{\em PTEP}
  {\bfseries 2015} no.~6, (2015) 061E01},
\href{http://arxiv.org/abs/1407.2030}{{\ttfamily arXiv:1407.2030 [gr-qc]}}.
%%CITATION = ARXIV:1407.2030;%%.

\bibitem{Sun:2019mqb}
L.~Sun, R.~Brito, and M.~Isi, ``{Search for ultralight bosons in Cygnus X-1
  with Advanced LIGO},''
\href{http://arxiv.org/abs/1909.11267}{{\ttfamily arXiv:1909.11267 [gr-qc]}}.
%%CITATION = ARXIV:1909.11267;%%.

\bibitem{Abuter:2018drb}
{\bfseries GRAVITY} Collaboration, R.~Abuter {\em et~al.}, ``{Detection of the
  gravitational redshift in the orbit of the star S2 near the Galactic centre
  massive black hole},''
  \href{http://dx.doi.org/10.1051/0004-6361/201833718}{{\em Astron. Astrophys.}
  {\bfseries 615} (2018) L15},
\href{http://arxiv.org/abs/1807.09409}{{\ttfamily arXiv:1807.09409
  [astro-ph.GA]}}.
%%CITATION = ARXIV:1807.09409;%%.

\bibitem{Naoz:2019sjx}
S.~Naoz, C.~M. Will, E.~Ramirez-Ruiz, A.~Hees, A.~M. Ghez, and T.~Do, ``{A
  hidden friend for the galactic center black hole, Sgr A*},''
\href{http://arxiv.org/abs/1912.04910}{{\ttfamily arXiv:1912.04910
  [astro-ph.GA]}}.
%%CITATION = ARXIV:1912.04910;%%.

\bibitem{Herdeiro:2014goa}
C.~A.~R. Herdeiro and E.~Radu, ``{Kerr black holes with scalar hair},''
  \href{http://dx.doi.org/10.1103/PhysRevLett.112.221101}{{\em Phys. Rev.
  Lett.} {\bfseries 112} (2014) 221101},
\href{http://arxiv.org/abs/1403.2757}{{\ttfamily arXiv:1403.2757 [gr-qc]}}.
%%CITATION = ARXIV:1403.2757;%%.

\bibitem{Binnington:2009bb}
T.~Binnington and E.~Poisson, ``{Relativistic theory of tidal Love numbers},''
  \href{http://dx.doi.org/10.1103/PhysRevD.80.084018}{{\em Phys. Rev.}
  {\bfseries D80} (2009) 084018},
\href{http://arxiv.org/abs/0906.1366}{{\ttfamily arXiv:0906.1366 [gr-qc]}}.
%%CITATION = ARXIV:0906.1366;%%.

\bibitem{Damour:2009vw}
T.~Damour and A.~Nagar, ``{Relativistic tidal properties of neutron stars},''
  \href{http://dx.doi.org/10.1103/PhysRevD.80.084035}{{\em Phys. Rev.}
  {\bfseries D80} (2009) 084035},
  \href{http://arxiv.org/abs/0906.0096}{{\ttfamily arXiv:0906.0096 [gr-qc]}}.

\bibitem{poisson_will_2014}
E.~Poisson and C.~M. Will,
  \href{http://dx.doi.org/10.1017/CBO9781139507486}{{\em Gravity: Newtonian,
  Post-Newtonian, Relativistic}}.
\newblock Cambridge University Press, 2014.

\bibitem{Page:2003rd}
D.~N. Page, ``{Classical and quantum decay of oscillatons: Oscillating
  selfgravitating real scalar field solitons},''
  \href{http://dx.doi.org/10.1103/PhysRevD.70.023002}{{\em Phys. Rev.}
  {\bfseries D70} (2004) 023002},
\href{http://arxiv.org/abs/gr-qc/0310006}{{\ttfamily arXiv:gr-qc/0310006
  [gr-qc]}}.
%%CITATION = GR-QC/0310006;%%.

\bibitem{Mendes:2016vdr}
R.~F.~P. Mendes and H.~Yang, ``{Tidal deformability of boson stars and dark
  matter clumps},'' \href{http://dx.doi.org/10.1088/1361-6382/aa842d}{{\em
  Class. Quant. Grav.} {\bfseries 34} no.~18, (2017) 185001},
\href{http://arxiv.org/abs/1606.03035}{{\ttfamily arXiv:1606.03035
  [astro-ph.CO]}}.
%%CITATION = ARXIV:1606.03035;%%.

\bibitem{Detweiler:1980uk}
S.~L. Detweiler, ``{KLEIN-GORDON EQUATION AND ROTATING BLACK HOLES},''
\href{http://dx.doi.org/10.1103/PhysRevD.22.2323}{{\em Phys. Rev.} {\bfseries
  D22} (1980) 2323--2326}.
%%CITATION = PHRVA,D22,2323;%%.

\bibitem{Baumann:2019eav}
D.~Baumann, H.~S. Chia, J.~Stout, and L.~ter Haar, ``{The Spectra of
  Gravitational Atoms},''
\href{http://arxiv.org/abs/1908.10370}{{\ttfamily arXiv:1908.10370 [gr-qc]}}.
%%CITATION = ARXIV:1908.10370;%%.

\bibitem{griffiths2017introduction}
D.~Griffiths, {\em Introduction to Quantum Mechanics}.
\newblock Cambridge University Press, 2017.
\newblock \url{https://books.google.pt/books?id=0h-nDAAAQBAJ}.

\bibitem{sakurai2011modern}
J.~Sakurai and J.~Napolitano, {\em Modern Quantum Mechanics}.
\newblock Addison Wesley, 2011.
\newblock \url{https://books.google.pt/books?id=N4I-AQAACAAJ}.

\end{thebibliography}\endgroup


\providecommand{\href}[2]{#2}\begingroup\raggedright\endgroup
\end{document}